\documentclass[twocolumn,epsfig,superscriptaddress,aps,prb]{revtex4-2}\usepackage[colorlinks=true,linkcolor=blue,anchorcolor=red,citecolor=blue, urlcolor=blue]{hyperref}
\usepackage{graphicx}
\usepackage{amssymb}
\usepackage{amsmath}
\usepackage{changes}

\begin{document}
\title{Conductance oscillations of antiferromagnetic layer tunnel junctions}
% \title{Conductance oscillations of magnetic topological junctions at a in-plane magnetic field}

\author{Sang-Jun Choi}
\email{sang-jun.choi@physik.uni-wuerzburg.de}
\affiliation{Institute for Theoretical Physics and Astrophysics, University of W\"urzburg, D-97074 W\"urzburg, Germany}
\author{Hai-Peng Sun}
\email{haipeng.sun@physik.uni-wuerzburg.de}
\affiliation{Institute for Theoretical Physics and Astrophysics, University of W\"urzburg, D-97074 W\"urzburg, Germany}
\author{Bj\"orn Trauzettel}
%\email{trauzettel@physik.uni-wuerzburg.de}
\affiliation{Institute for Theoretical Physics and Astrophysics, University of W\"urzburg, D-97074 W\"urzburg, Germany}
\affiliation{W\"urzburg-Dresden Cluster of Excellence ct.qmat, Germany}
\date{\today}

\begin{abstract}
{
We study conductance oscillations of antiferromagnetic layer tunnel junctions composed of antiferromagnetic topological insulators (MTIs) such as MnBi$_{2}$Te$_{4}$. In presence of an in-plane magnetic field, we find that the two terminal differential conductance across the junction oscillates as a function of field strength. Notably, the quantum interference at weak fields for the even-layer case is distinctive from the odd-layer case due to the scattering phase shift $\pi$. Consequently, the differential conductance is vanishing (maximized) at integer magnetic flux quanta for even-layer (odd-layer) junctions. The conductance oscillations manifest the layer-dependent quantum interference in which symmetries and scattering phases play essential roles. In numerical calculations, we observe that the quantum interference undergoes an evolution from SQUID-like to Fraunhofer-like oscillations as the junction length increases.

}
\end{abstract}
\maketitle

\section{Introduction}
Magnetism and topology are two main research areas of modern physics. Many efforts have been devoted to study the interplay of magnetism and topology in magnetic topological materials, for instance, in magnetic topological insulators~\cite{Liu16ARCMPQuantum,Tokura19NRPMagnetic,Wang21IIntrinsic,Bernevig22NProgress,Chang23RMPColloquium,Wang23NSRTopological,Li23NSRProgress}. The quantum anomalous Hall effect is one of the key features of magnetic topological insulators that can be employed in prototype quantum devices~\cite{He14NSRQuantum,Bhattacharyya21AMRecent,Chi22AMProgress}. However, the working temperature of the quantum anomalous Hall effect is limited by the small surface gap and disorder in magnetically doped topological insulators. The recently discovered intrinsic magnetic topological insulator MnBi$_{2}$Te$_{4}$ gives hope for the observation of the quantum anomalous Hall effect at high temperatures~\cite{Otrokov19NPrediction,Deng20SQuantum,Deng21NPHightemperature}. It is a versatile material with rich topological physics~\cite{Li19SAIntrinsic,Li19PRXDirac,Hao19PRXGapless,Zhang20PRLMobius,Sun20PRBAnalytical,Wu20PRXDistinct,Yang21PRXOddEven,Li21PRLCritical,Chen23PRBSidesurfacemediated}, such as axion insulators~\cite{Zhang19PRLTopological, Otrokov19PRLUnique,Liu20NMRobust,Gao21NLayer}, higher Chern numbers~\cite{Ge20NSRHighChernnumber} and type-II Weyl semimetals~\cite{Lee21PRXEvidence,Lei22PRBMagnetically}. 

In thin films of magnetic topological insulators, top and bottom surface states hybridize due to finite-size effects. Previous studies have shown that hybridization can be exploited to tune the properties of topological insulators, which can be utilized in prototype topological electronics~\cite{Xu19PRLTuning,Chong22NCEmergent}, especially in topological transistors~\cite{Michetti13APLDevices,Qian14SQuantum,Liu14NMSpinfiltered,Wang15PRLElectrically,Liu15NLSwitching,Collins18NElectricfieldtuned}.  Recently, a new mechanism of magnetic topological transistors has been proposed based on the layer degree of freedom of MnBi$_{2}$Te$_{4}$~\cite{Sun23PRRMagnetic}. The magnetic topological transistor works well for even-layer thin films thanks to compensated magnetization, whereas hybridization hampers the performance in odd-layer thin films. Up to now, the influence of in-plane magnetic fields on the performance of magnetic topological transistors has not been analyzed. We show it in this work that magnetic fields substantially affect the transport characteristics of such devices. 

To do so, we investigate the scattering processes of an antiferromagnetic layer tunnel junction (ALTJ) composed of intrinsic magnetic topological insulators MnBi$_{2}$Te$_{4}$ in presence of hybridization. The junction is formed between two regions with (anti-)parallel electric fields, applied perpendicularly to the layers of the magnetic topological insulators. We provide analytical results of the scattering matrices and the transmission probability for both odd- and even-layer junctions with and without in-plane magnetic fields. In absence of magnetic fields, quantum interference is constructive for the odd-layer case, while it is destructive for the even-layer case due to symmetry constraints. In presence of in-plane magnetic fields, the two terminal differential conductance oscillates as the magnetic field increases. The conductance oscillations manifest the layer-dependent quantum interference. We observe that this quantum interference undergoes an evolution from SQUID-like to Fraunhofer-like oscillations as the junction length increases. These features indicate particular quantum interference in ALTJs that allows us to observe characteristic properties of topological surface states.

This article is organized as follows. In Sec.~\ref{Sec:Model}, we provide the effective surface model for intrinsic antiferromagnetic topological insulators MnBi$_{2}$Te$_{4}$ in presence of vertical electric fields and in-plane magnetic fields. We also describe the method used for numerical simulations. In Sec.~\ref{Sec:T(E)}, we show the analytical results of the scattering matrix and transmission probability of ALTJs for both odd- and even-layer films. The transport results of ALTJs in presence of in-plane magnetic fields are presented in Sec.~\ref{Sec:OscG}. We conclude in Sec.~\ref{Sec:discussion}.

\section{Model}\label{Sec:Model}
We provide theoretical models which are used to calculate the conductance oscillations of ALTJs. In Sec.~\ref{Sec:H}, we present the low-energy effective Hamiltonian. In Sec.~\ref{Sec:Landauer}, we introduce the theoretical formalism for the calculation of the two terminal electrical conductance.

\begin{figure}[t] 
\centering
\includegraphics[width=0.88\columnwidth]{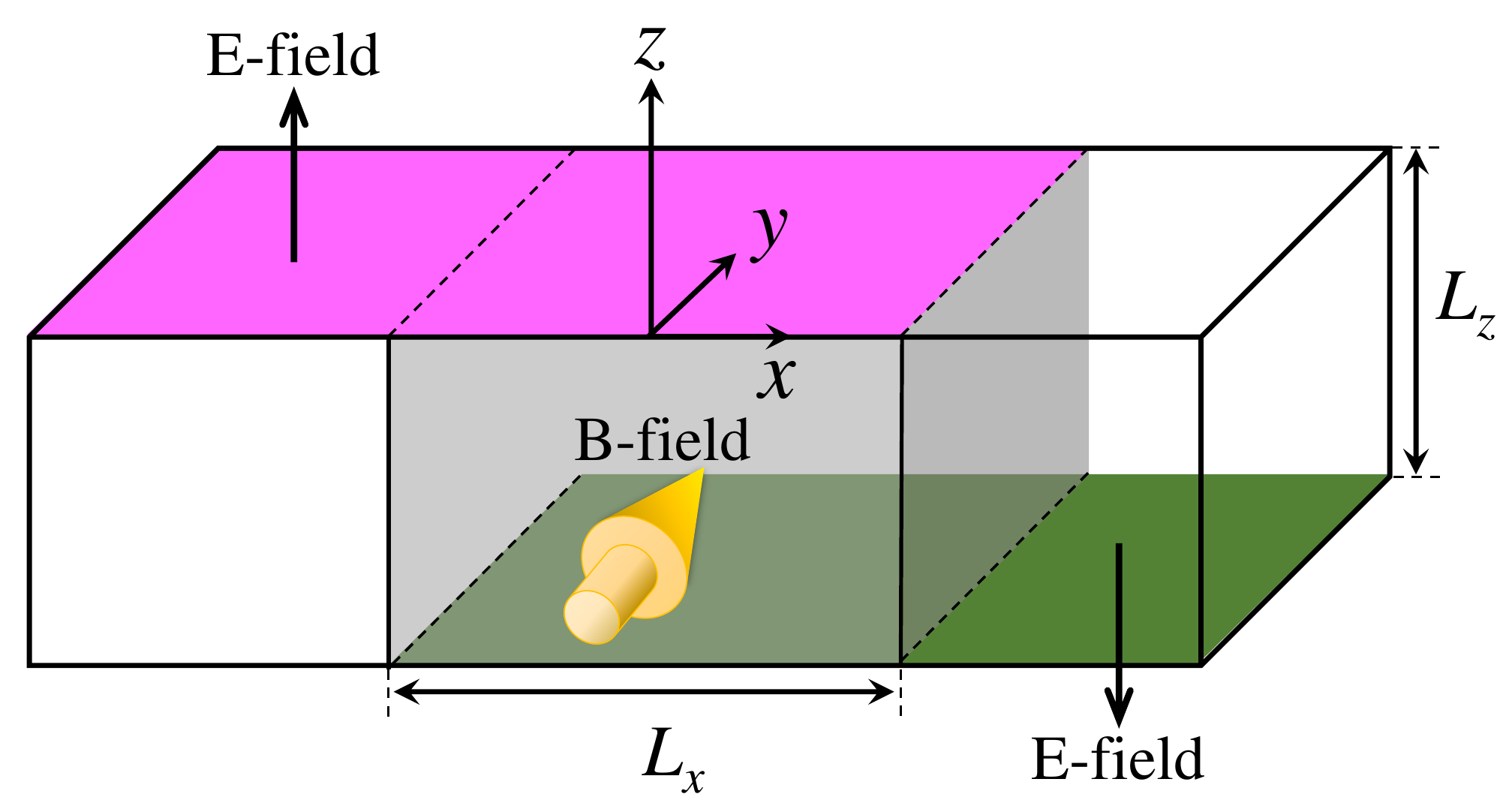}\\
\caption{
Schematic view of the ALTJ. Out-of-plane electric fields are applied to the left ($x<-L_x/2$) and the right ($x>L_x/2$) regions. The junction region is assumed to be free of external electric fields. The purple and green colors depict top and bottom surface states crossing the Fermi level, respectively. An in-plane magnetic field is applied along the $y$-axis.
}\label{Fig1}
\end{figure}

\subsection{Low-energy effective Hamiltonian} \label{Sec:H}
We begin with an effective model of antiferromagnetic MnBi$_2$Te$_4$ thin films in presence of out-of-plane electric fields and in-plane magnetic fields. MnBi$_2$Te$_4$ is a magnetic topological insulator with intrinsic antiferromagnetic order~\cite{Zhang19PRLTopological,Otrokov19NPrediction,Gong19CPLExperimental}.  
Out-of-plane electric fields can be realized by dual gates~\cite{Gao21NLayer,Cai22NCElectric}. We assume that the resulting electric potential is an odd function of $z$ without loss of generality. To derive the effective surface model, we project the bulk Hamiltonian onto the surface states of the topological insulator~\cite{Lu10PRBMassive,Shan10NJPEffective,Sun20PRBAnalytical}. Thus, the effective surface model reads~\cite{Sun23PRRMagnetic}
\begin{align}
	H(x,y) = H_0 + H_{V} + H_{ex},
	\label{Eq:H}
\end{align}
where $H_0$ describes the Dirac surface states of topological insulators given by
\begin{align} \label{Eq:Effective_H0}
	H_0(x,y) = \tau_z \otimes [v_F(-i\hbar\nabla - e\vec{A})\times\vec{\sigma}]_z + m_{tb}\tau_x\otimes\sigma_0.
\end{align}
The Pauli matrices $\vec{\tau}$ and $\vec{\sigma}$ act on top and bottom surface state and spin space, respectively. The basis of the model is $\{|t,\uparrow\rangle, |t,\downarrow\rangle, |b,\uparrow\rangle, |b,\downarrow\rangle\}$, where $|t,\sigma\rangle$ and $|b,\sigma\rangle$ are the lowest-energy eigenstates of topological insulators at the $\Gamma$ point with spin $\sigma=\uparrow,\downarrow$. We consider the in-plane magnetic field with minimal coupling to a vector potential $\vec{A}$ describing a magnetic field along the $y$-axis, i.e., $\nabla\times\vec{A}=B\hat{e}_y$. The last term in Eq.~\eqref{Eq:Effective_H0} describes the overlap between top and bottom surface states with coupling strength $m_{tb} = M_0 e^{-\lambda L_z}$, which is exponentially small for thick films. $L_z$ is the thickness of the MTI film. $\lambda$ is the localization length of the surface states along the $z$-axis. $M_0$ is a bulk parameter of the material MnBi$_2$Te$_4$. Both $\lambda$ and $M_0$ are determined from parameter sets of MnBi$_2$Te$_4$ based on {\it ab initio} calculations~\cite{Zhang19PRLTopological}.

An ALTJ is formed between the regions with two  electric fields [see Fig.~\ref{Fig1}], described by $H_V$ in Eq.~\eqref{Eq:H},
\begin{eqnarray}
	H_{V}(x) = V_g(x) \tau_z \otimes \sigma_0, \label{Eq:Vfield}
\end{eqnarray}
where $V_g(x) = V_g\Theta(-x-L_x/2) - V_g\Theta(x-L_x/2)$ is the effective electric potential acting on the surface states with $V_g$ being the strength of the gate potential. The magnetic junction is formed in the region $-L_x/2<x<L_x/2$ with $V_g(x)=0$. The effective electric potential acting on the left (right) region of the junction is $V_g$ ($-V_g$). Hence, $H_V$ shifts top and bottom surface states in energy oppositely~\cite{Sun23PRRMagnetic}.

The last term $H_{ex}$ in Eq.~\eqref{Eq:H} describes the effective exchange field of the surface states. Due to the antiferromagnetic order in the bulk, the form of $H_{ex}$ is layer number dependent. For odd-layer films, the directions of the effective exchange fields at top and bottom surfaces are the same, leading to
\begin{eqnarray}
	H_{ex}^\text{odd} = m_\text{o} \tau_0 \otimes \sigma_z.
\end{eqnarray}
For even-layer films, the directions of the effective exchange fields at top and bottom surfaces are opposite, leading to
\begin{eqnarray}
	H_{ex}^\text{even} = m_\text{e} \tau_z \otimes \sigma_z.
\end{eqnarray}
$m_\text{o}$ and $m_\text{e}$ are the strength of the corresponding effective exchange field~\cite{Sun23PRRMagnetic}.

\begin{figure}[b] 
\centering
\includegraphics[width=0.99\columnwidth]{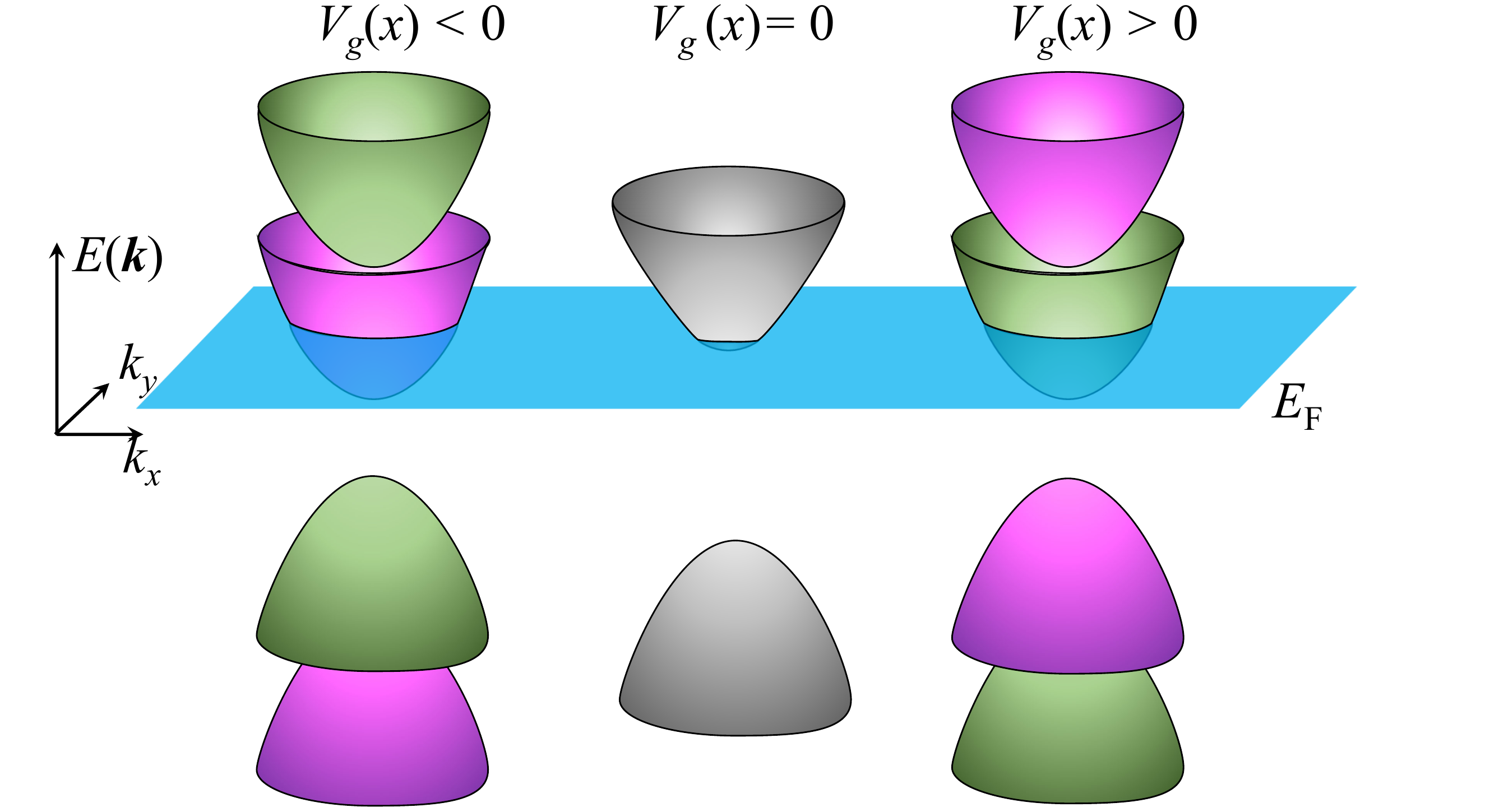}
\caption{
Band alignment of the ALTJ for the calculation of the transmission probability. In the left region, where $V_g(x)<0$, only the top surface state (purple) crosses the Fermi energy $E_F$. In the right region, where $V_g(x)>0$, the bottom surface state (green) crosses the Fermi energy. In the junction region, top and bottom surface states are nearly degenerate such that both cross the Fermi energy.
}\label{Fig2}
\end{figure}

\subsection{Two terminal conductance} \label{Sec:Landauer}
We employ the Landauer formalism to evaluate the two terminal electrical conductance~\cite{Landauer70Electrical,Buttiker86Four,Datta97Electronic}.
\begin{equation}
I = G_0\int dE [f(E-eV)-f(E)]T(E),
\end{equation}
where $G_0=e^2/h$ is the conductance quantum, $f(E)$ is the Fermi distribution function. $T(E)$ is the transmission probability from the left region $x<-L_x/2$ to the right region $x>L_x/2$ (see Fig.~\ref{Fig2}). We obtain the differential conductance $G=dI/dV$ as $G=G_0T(E_F)$ at zero temperature, where $E_F$ is Fermi energy.

We calculate the transmission probability $T(E)$ analytically for short junctions in Sec.~\ref{Sec:T(E)} using the effective Hamiltonian of Eq.~\eqref{Eq:H}. In addition, we complement the analytical calculation with numerical calculation of $T(E)$ 
for long junctions in Sec.~\ref{Sec:OscG}. We conduct the numerical simulation considering the discretized 3D bulk Hamiltonian $H_\text{cubic}$ on a cubic lattice and exploiting the recursive Green function technique, where
\begin{equation}
	T(E)= \mathrm{tr}[\Gamma_{L} G^r\Gamma_R G^a].
\end{equation}
Here, $G^r=[E-(H_\text{cubic}+\Sigma_L+\Sigma_R)]^{-1}$ is the retarded Green function of the junction region, $-L_x/2<x<L_x/2$.  $\Sigma_{L}$ and $\Sigma_{R}$ are the self-energies of the left and right regions, respectively. $\Gamma_{L/R}=i(\Sigma_{L/R}^r-\Sigma_{L/R}^a)$ are the linewidth functions. $G^a$ is the advanced Green function.

\section{Transmission probability $T$ without a magnetic field} \label{Sec:T(E)}
To reveal the physical origin for the conductance oscillations in presence of an in-plane magnetic field, we analyze the transmission probability in terms of the physical scattering processes without magnetic field first. Interestingly, we identify scattering processes involving distinct quantum interferences. We show that the quantum interference for odd- and even-layer films, respectively, occurs constructively and destructively due to particular symmetries between top and bottom layers. In Sec.~\ref{Sec:Smatrices}, we present the generic scattering processes. In Sec.~\ref{Sec:T(E)Odd} and \ref{Sec:T(E)Even}, we provide analytical results of the transmission probability $T$ for odd- and even-layer ALTJs demonstrating the constructive and destructive interference.

We obtain analytical results of  $T$ using the effective Hamiltonian Eq.~\eqref{Eq:H}. To simplify the analysis, we focus on the case that the Fermi energy is close to the band bottom of the junction region, $m_{\text{o,e}}\lesssim E_F$. In this regime, $T$ is dominated by the transmission of electrons with perpendicular angle of incidence at the junction. Then, the scattering problem effectively becomes a one-dimensional scattering problem between top and bottom layers along the $x$-axis. 

\begin{figure}[b] 
\centering
\includegraphics[width=0.99\columnwidth]{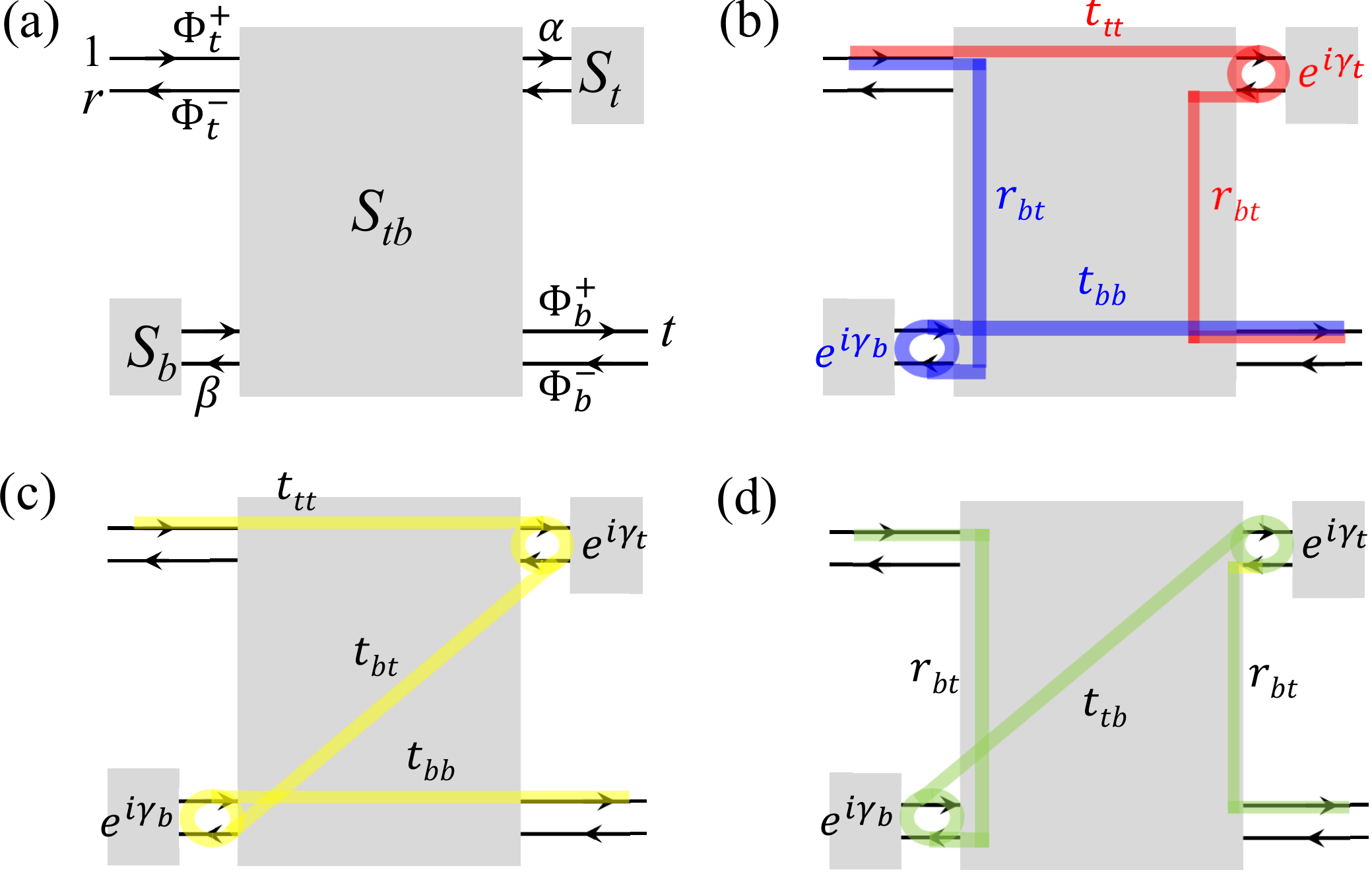}
\caption{
(a) By combining scattering matrices $S_t$, $S_b$, and $S_{tb}$, we calculate the transmission probability $T=|t|^2$. $\Phi^{+/-}_t$ ($\Phi^{+/-}_b$) are the right/left going modes of top (bottom) surface states. (b-d) Possible scattering processes for thick films. (b) Most dominating scattering processes for weak connection between top and bottom layers. Note that two scattering processes in (b) form a quantum interference loop enclosing a magnetic flux (if present). (c) Next relevant scattering process for thick layers. (d) Third order scattering between top and bottom layers, typically negligible compared to (b) and (c).
}\label{Fig3}
\end{figure}

\subsection{Scattering matrices} \label{Sec:Smatrices}
We now describe the scattering matrices $S_t$, $S_b$, and $S_{tb}$ used to calculate $T$ (see Fig.~\ref{Fig3}). $S_{b}$ ($S_{t}$) is the scattering matrix of bottom (top) surface state in the left (right) region. As the gates deplete one of the surface states, the incident electrons in each case are fully reflected accumulating the phase factor $S_{t(b)}=e^{i\gamma_{t(b)}}$  (see Fig.~\ref{Fig2} and Fig.~\ref{Fig3}). Meanwhile, in the junction region, both  top and bottom surface states are filled up to the Fermi energy. We introduce $S_{tb}$ to describe electron scattering in the junction region,
\begin{equation}
S_{tb} = \left(
\begin{array}{cccc}
t_{tt}  &  r_{tt}  &  t_{tb}  &  r_{tb} \\
r_{tt}  &  t_{tt}  &  r_{tb}  &  t_{tb} \\
t_{bt}  &  r_{bt}  &  t_{bb}  &  r_{bb} \\
r_{bt}  &  t_{bt}  &  r_{bb}  &  t_{bb}
\end{array}\right),
\end{equation}
where $S_{tb}(\Phi_t^+,\, \Phi_t^-,\, \Phi_b^+,\, \Phi_b^-)^T_\text{in}=(\Phi_t^+,\, \Phi_t^-,\, \Phi_b^+,\, \Phi_b^-)^T_\text{out}$. 

We derive the transmission amplitude $t$ by combining the scattering matrices $S_t$, $S_b$, and $S_{tb}$,
\begin{equation}
S_{tb}\left(
\begin{array}{c}
1 \\
\alpha e^{i\gamma_t} \\
\beta e^{i\gamma_b} \\
0
\end{array}\right)
=\left(
\begin{array}{c}
\alpha \\
r \\
t \\
\beta
\end{array}\right),
\end{equation}
where $\alpha, r, t, \beta$ are outgoing scattering amplitudes [see Fig.~\ref{Fig3}(a)]. The resulting transmission amplitude is
\begin{eqnarray}
t = t_{bt} &+& \frac{ r_{bt}\rho_{tt}t_{tt} + t_{bb}\rho_{bb}r_{bt} }{ 1 - \rho_{tt}t_{tb}\rho_{bb}t_{bt} } \nonumber\\
 &+& \frac{ t_{bb}\rho_{bb}t_{bt}\rho_{tt}t_{tt}}{ 1 - \rho_{tt}t_{tb}\rho_{bb}t_{bt} } + \frac{ r_{bt}\rho_{tt}t_{tb}\rho_{bb}r_{bt} }{ 1 - \rho_{tt}t_{tb}\rho_{bb}t_{bt} }, \label{Eq:t}
\end{eqnarray}
where $\rho_{tt} = e^{i\gamma_t}/(1-r_{tt}e^{i\gamma_t})$ appears due to multiple phase accumulations at the top layer of the right region. $\rho_{bb} = e^{i\gamma_b}/(1-r_{bb}e^{i\gamma_b})$ occurs similarly at the bottom layer of the left region. For thick ALTJs, the denominators of second, third, and fourth terms can be approximated by $1/(1 - \rho_t \rho_b t_{tb}t_{bt})\approx 1$, since higher orders of the scattering between top and bottom layers are negligibly small for thick layers. The second term in Eq.(\ref{Eq:t}) corresponds to the scattering processes illustrated in Fig.~\ref{Fig3}(b), involving (flux-dependent) quantum interference. The third and fourth terms correspond to those illustrated in Figs.~\ref{Fig3}(c) and~\ref{Fig3}(d), respectively.

\begin{figure}[b] 
\centering
\includegraphics[width=0.99\columnwidth]{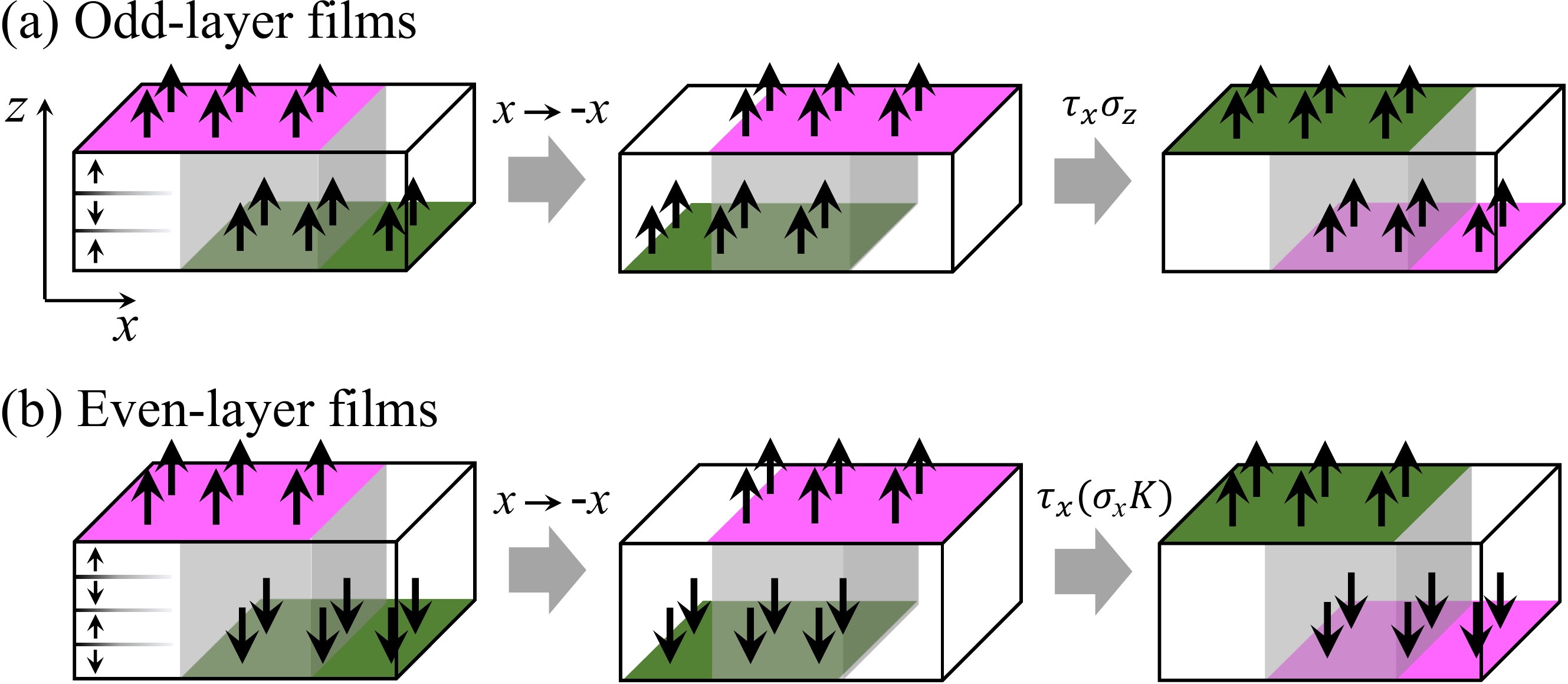}
\caption{Symmetries of ALTJs due to the antiferromagnetic ordering across the layers of ALTJs. The up and down arrows represent the direction of magnetization.  In (a), the odd-layer films exhibits space inversion symmetry between top and bottom layers of ALTJs. In (b), ALTJs of even-layer films acquires a particular type of inversion symmetry with space inversion and additional spin inversion.
}\label{Fig:Sym}
\end{figure}

\subsection{Transmission through odd-layer junctions} \label{Sec:T(E)Odd}
We show that the inversion symmetry of ALTJs of odd-layer films results in constructive quantum interference. We begin with providing the scattering matrix for odd-layer films:
\begin{equation}
S_{tb}^\text{odd} = \left(
\begin{array}{cccc}
t'_\text{o}  &  r'_\text{o}  &  t''_\text{o}  &  r''_\text{o} \\
r'_\text{o}  &  t'_\text{o}  &  r''_\text{o}  &  t''_\text{o} \\
t''_\text{o}  &  r''_\text{o}  &  t'_\text{o}  &  r'_\text{o} \\
r''_\text{o}  &  t''_\text{o}  &  r'_\text{o}  &  t'_\text{o}
\end{array}\right),
\end{equation}
where $t_{tt}=t_{bb}=t'_\text{o}$, $r_{tt}=r_{bb}=r'_\text{o}$, $t_{bt}=t_{tb}=t''_\text{o}$, and $r_{bt}=r_{tb}=r''_\text{o}$ due to inversion symmetry $\tau_x\sigma_z H(-x) \tau_x\sigma_z = H(x)$ [see Fig.~\ref{Fig:Sym}(a)]. Hence, we can simplify the transmission amplitude in Eq.~\eqref{Eq:t} for odd-layer films as
\begin{equation}
t_\text{o} = t_\text{o}'' + r_\text{o}''t_\text{o}'\left(e^{i\gamma_t}+e^{i\gamma_b}\right), \label{Eq:todd}
\end{equation}
where scattering processes are kept up to the second order of the amplitudes $t'_\text{o}$, $r'_\text{o}$, $t''_\text{o}$, and $r''_\text{o}$ to focus on the dominating ones [Derivations of $t_\text{o}'$, $r_\text{o}'$, $t_\text{o}''$, and $r_\text{o}''$ are provided in Appendix~\ref{Sec:Appendix}]. We note that the second term in Eq.~\eqref{Eq:todd} is responsible for quantum interference.

Moreover, we identify that the inversion symmetry of ALTJs of odd-layer films results in constructive quantum interference at zero magnetic field. The scattering phase $\gamma_t$ in Eq.~\eqref{Eq:todd} can be written as
\begin{equation}
\gamma_t = \text{arg}\left\{\frac{E_F+V_g-i\sqrt{m_\text{o}^2-(E_F+V_g)^2}}{m_\text{o}}\right\}. \label{Eq:gammat}
\end{equation}
The value of $\gamma_t$ alone depends on various parameters such as Fermi energy $E_F$, gate voltage $V_g$, and effective magnetization at the surface $m_\text{o}$. However, we find that the difference of the phases $\gamma_t$ and $\gamma_b$ vanishes owing to inversion symmetry, resulting in constructive quantum interference.

\subsection{Transmission through even-layer junctions} \label{Sec:T(E)Even}
Due to the antiferromagnetic ordering across the layers of ALTJs, the transmission $t$ of even-layer films shows distinct features from odd-layer films [see Fig.~\ref{Fig:Sym}]. We show that the particular type of the inversion symmetry of even-layer films, $\tau_x\otimes\sigma_x H^*(-x) \tau_x\otimes\sigma_x = H(x)$  followed by spin inversion, results in destructive quantum interference for the transmission probability.

From the particular inversion symmetry of even-layer films, we find that the transmission between top and bottom layer are prohibited $t_{bt}=t_{tb}=0$. This is because spin-flip tunneling between top and bottom layers is blocked. Hence, only the scattering process shown in Fig.~\ref{Fig3}(b) is responsible for the transmission in even-layer films. The transmission amplitude $t_\text{e}$ for even-layer films becomes
\begin{equation}
t_\text{e} = r_\text{e}''t_\text{e}'\left(e^{i\gamma_t}+e^{i\gamma_b}\right), \label{Eq:teven}
\end{equation}
where all orders of scattering processes are kept. The scattering matrix for even-layer films is given by
\begin{equation}
S_{tb}^\text{even} = \left(
\begin{array}{cccc}
t'_\text{e}  &  r'_\text{e}  &  0  &  r''_\text{e} \\
r'_\text{e}  &  t'_\text{e}  &  r''_\text{e}  &  0 \\
0  &  r''_\text{e}  &  t'_\text{e}  &  -r'_\text{e} \\
r''_\text{e}  &  0  &  -r'_\text{e}  &  t'_\text{e}
\end{array}\right),
\end{equation}
where $t_{tt}=t_{bb}=t'_\text{e}$ and $r_{bt}=r_{tb}=r''_\text{e}$. [Matrix elements of $S_{tb}^\text{even}$ are provided in Appendix~\ref{Sec:Appendix}.]

While the scattering phase $\gamma_t$ is again the one in Eq.~\eqref{Eq:gammat}, we find that the particular inversion symmetry of even-layer films enforces $\gamma_t^\text{even}-\gamma_b^\text{even}= \pi$. Hence, the quantum interference of the ALTJ of even-layer films is destructive with $t_\text{e}=0$. We note that the scattering phase shift $\pi$ stems from the geometric phase associated with topological junctions~\cite{13Choi}.

\section{Conductance oscillations due to in-plane magnetic fields} \label{Sec:OscG}
We study conductance oscillations $G(B)/G_0=T(B)$ of ALTJs in presence of in-plane magnetic fields $B$. 
First, we evaluate $T(B)$ analytically using the effective Hamiltonian Eq.~\eqref{Eq:H} and analyze the relevant scattering processes in Sec.\ref{Sec:flux_Ana}. The analytical results show that the conductance oscillations $T(B)$ of odd-layer junctions are maximized at integer magnetic flux quanta. In contrast, we identify that $T(B)$ for even-layer junctions is shifted by the half magnetic flux quantum due to the scattering phase difference $\gamma_t^\text{even}-\gamma_b^\text{even}= \pi$. Second, we provide numerical results of the conductance oscillations for various lengths in Sec.~\ref{Sec:flux_Num}. Interestingly, we find that the conductance oscillations exhibit a crossover from SQUID-like to Fraunhofer-like behavior as the length of the junction increases.

\begin{figure}[b] 
\centering
\includegraphics[width=0.99\columnwidth]{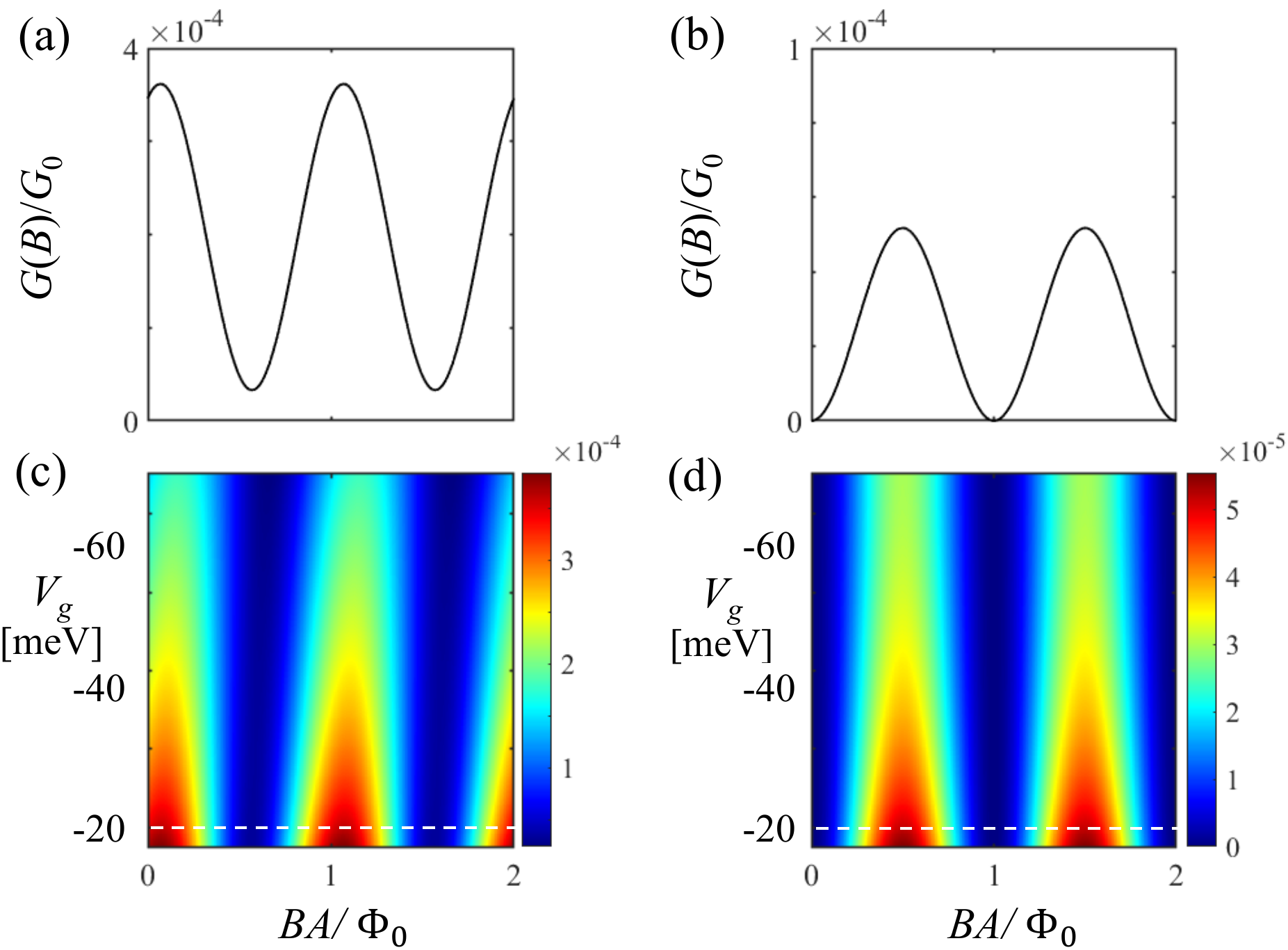}
\caption{Conductance oscillations $G(B)/G_0$ of short ALTJs of odd-layer films (left panels) and even-layer films (right panels). The conductance oscillations of odd- and even-layer films at $V_g=-20\,\text{meV}$ are depicted in (a) and (b), respectively. The density plots of $G(B)/G_0$ are provided in (c) and (d) for odd- and even-layer films, respectively. We use parameters $E_F=36\,\text{meV}$, $m_\text{e}=m_\text{o}=35.02\,\text{meV}$, and $m_{tb} = 0.35\,\text{meV}$ for 11 and 10 quintuple layers. The length of the junction is $L_x=4\,\text{nm}$, which is much smaller than $\hbar v_F/E_F = 9\,\text{nm}$.
}\label{Fig4}
\end{figure}

\subsection{Analytical calculations for short junctions}\label{Sec:flux_Ana}
We obtain the conductance oscillations $G(B)$ from the transmission $T(B)=G(B)/G_0$ at a given Fermi energy $E_F$. We note that various scattering paths in the junction region accumulate different quantum phases from the in-plane magnetic field. For short junctions $L_x\ll \hbar v_F/E_F$, the dynamical phase is substantially smaller than the Peierl's phase of the magnetic field. We consider thick films for which higher orders of interlayer scattering are negligible. We take the magnetic field into account as phase shifts of the scattering matrix $S_{tb}$ for a given gauge choice,
\begin{equation}
S_{tb} \mapsto \left(
\begin{array}{cccc}
t_{tt}e^{i\nu}  &  r_{tt}  &  t_{tb}  &  r_{tb} \\
r_{tt}  &  t_{tt}e^{-i\nu}  &  r_{tb}  &  t_{tb} \\
t_{bt}  &  r_{bt}  &  t_{bb} &  r_{bb} \\
r_{bt}  &  t_{bt}  &  r_{bb}  &  t_{bb}
\end{array}\right),
\end{equation}
with $\nu = 2\pi BA/\Phi_0$, $A=L_xL_z$ the in-plane area of the junction, and $\Phi_0=h/e$ the magnetic flux quantum. We attribute the conductance oscillations to quantum interference. Note that the scattering processes shown in Fig.~\ref{Fig3}(b) dominate the transmission across the junction.

We first show that conductance oscillations $T(B)$ for the odd-layer case as a function of magnetic field [Figs.~\ref{Fig4}(a) and (c)]. Unfortunately, the transmission probability for the odd-layer case does not allow a simple expression. However, we provide an approximated solution for odd-layer case in the leading order of interlayer scattering $t''_\text{o}$ and $r''_\text{o}$ as
\begin{align}
& T^\text{odd}(B)  \\
& =  \left| r''_\text{o}\frac{(1+e^{i\nu})t'_\text{o}}{1-r'_\text{o}e^{i\gamma}} + t''_\text{o} \left[ e^{-i\gamma_t} + e^{i\nu}\left( \frac{t'_\text{o} e^{i\gamma_t}}{1-r'_\text{o}e^{i\gamma}} \right)^2 \right] \right|^2, \nonumber
\end{align}
where $\nu = 2\pi BA/\Phi_0$. The transmission probability of the odd-layer case is approximately reduced to the numerator of the first term $1+e^{i\nu}$. Accordingly, $T^\text{odd}(B)$ is maximized at integer multiples of magnetic flux quanta. 

We now present $T(B)$ for the even-layer case [Figs.~\ref{Fig4}(b) and (d)]. It can understood by the analytical result of the transmission probability
\begin{equation}
T^\text{even}(B) = \frac{ 2k_F^2\sin^2(k_FL_x)[1+\cos(2\pi BA/\Phi_0+\Delta\gamma)]}{ C(V_g)D(V_g) },
\end{equation}
where $k_F = \sqrt{E_F^2 - m_\text{e}^2 - m_{tb}^2}/(\hbar v_F)$,
$C(V_g)=[\sqrt{m_\text{e}^2-(E_F+V_g)^2}\sin(k_FL_x) + \hbar v_F k_F\cos(k_FL_x) ]^2 + V_g^2\sin^2(k_FL_x)$, 
$D(V_g)= [(m_\text{e}^2 + m_{tb}^2)\sin^2(k_FL_x) + (\hbar v_F k_F)^2]/(\hbar v_F m_{tb})^2$,
and $\Delta\gamma=\gamma_t-\gamma_b$.
Due to the scattering phase shift $\Delta\gamma=\pi$, the conductance oscillation $T^\text{even}(B)$ of the even-layer junctions is shifted by half the magnetic flux quantum from $T^\text{odd}(B)$ of the odd-layer case. Hence, the conductance oscillates as the magnetic field increases and $T^\text{even}(B)$ vanishes at integer multiples of magnetic flux quanta.

\begin{figure}[t] 
\centering
\includegraphics[width=0.99\columnwidth]{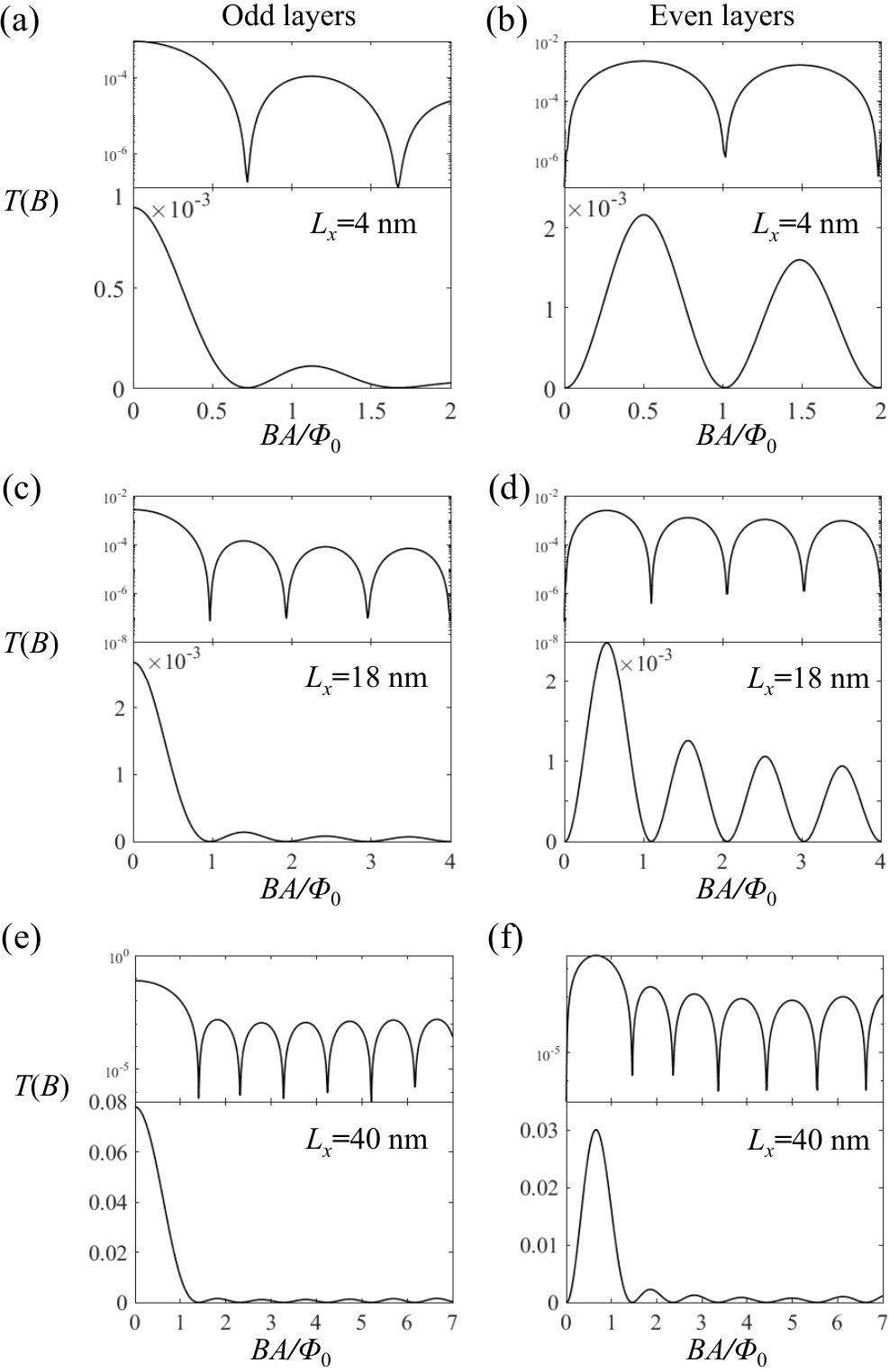}
\caption{Conductance oscillations $G/G_0=T(B)$ of ALTJs as a function of magnetic field $B$ for (a,c,e) odd and (b,d,f) even layers with different junction lengths $L_x = 4, 18, 40\,\text{nm}$. For better displays of the nodes $G/G_0\sim 0$, we provide $G/G_0$ with log scale as an upper inset of each panel.The numerical calculations are obtained with 11 and 10 layers with $E_F=36$ meV.
}\label{Fig5}
\end{figure}

\subsection{Numerical calculations for short and long junctions}\label{Sec:flux_Num}
The conductance oscillations $G(B)/G_0=T(B)$ of short ALTJs in Sec.~\ref{Sec:flux_Ana} require strong magnetic fields due to the small lateral cross section of short junctions. To overcome this limitation, we numerically study the conductance oscillations for various junction lengths in this section. 

For short junctions of odd-layer films, we find the numerically obtained conductance oscillation $T(B)$ shows qualitatively similar behaviors to the analytical result in Sec.~\ref{Sec:flux_Ana} in that $T(B)$ is maximized around the integer multiples of magnetic flux quanta [Fig.~\ref{Fig5}(a)]. However, the visibility of the numerically obtained $T(B)$ decreases fast as increasing magnetic fields $B$. This is because the numerical calculation includes more realistic scattering paths forming various interference loops enclosing smaller magnetic flux quanta than analytical calculations. On the other hand, the numerically obtained conductance oscillation $T(B)$ for even-layer films exhibits more similar behaviors to the analytical results in Sec.~\ref{Sec:flux_Ana} with vanishing $T(B)$ at the integer multiples of magnetic flux quanta but with the much slower decaying visibility. We attribute the slower decaying visibility of even-layer junctions to the fact that only the scattering processes in Fig.~\ref{Fig3}(b) are responsible for the conductance oscillation $T(B)$ of even-layer junctions, while all scattering processes in Fig.~\ref{Fig3} contribute to $T(B)$ of odd-layer junctions. Hence, when we focus on the visibilities, the analytical calculations for even-layer films becomes more consistent with realistic numerical calculations than that for the odd-layer films. However, we highlight that the overall oscillatory behaviors of analytical and numerical calculation exhibit the same results, when we focus on the magnetic flux quanta maximizing and minimizing $T(B)$.

For long junctions at high magnetic fields, the accumulated quantum phases of different scattering paths can vary significantly [Figs.~\ref{Fig5}(e) and (f)],  beyond the scope of the analytical method. We find that the various scattering paths for the conductance oscillations result in Fraunhofer-like interference patterns. Figs.~\ref{Fig5}(c) and (d) show intermediate junction lengths with SQUID-like [Figs.~\ref{Fig5}(a) and (b)] and Fraunhofer-like oscillation patterns [Figs.~\ref{Fig5}(e) and (f)]. However, for weak magnetic fields $|B|A<\Phi_0$, the quantum interferences for the even- and odd-layer junctions remain the same regardless of the length of junctions. For the even-layer case, it becomes finite from zero and oscillates as the magnetic field increases. For the odd-layer case, the conductance peak is reduced due to the presence of the in-plane magnetic field.

% It is opposite to the Josephson junction made of Dirac semimetals, where for short junction the quantum interference is Fraunhoffer oscillation and for long junction is SQUID-like oscillation.

% \subsection{Magnetic topological transistor}

% {\color{magenta}
% [Numerical results of (i) the regime corresponding to the analytical calculation for a fixed Fermi energy]
% [Numerical result of (ii) conductance map as a function of magnetic field and Fermi energy]
% [Note that the numerical calculation should be conducted with large $L_y$ to see the effect of the various incidence angles.]
% }

\section{Conclusion}\label{Sec:discussion}

To summarize, we study the scattering processes of an antiferromagnetic layer tunnel junction composed of intrinsic magnetic topological insulators MnBi$_{2}$Te$_{4}$ in presence of hybridization. 
We obtain analytical results of the
scattering matrices and the transmission probabilities for
both odd- and even-layer junctions with and without in-
plane magnetic fields. 
We show that the two paths of most dominating scattering processes at first order of the small scattering between top and bottom layers form a quantum interference loop enclosing a magnetic flux in presence of in-plane magnetic fields.
Importantly, we reveal that the scattering processes involving distinct quantum interference in odd- and even-layer films at weak fields are constructive and destructive, respectively, due to their particular symmetries. Thus, the differential conductance is vanishing (maximized) at integer magnetic flux quanta for even-layer (odd-layer) junctions.
Additionally, we observe that the quantum interference undergoes an evolution from SQUID-like to Fraunhofer-like oscillations as the junction length increases. It indicates that for a long ALTJ various scattering paths in the junction region get involved.  These features imply rich quantum interference in ALTJs that makes the observation of characteristic properties of topological surface states in ALTJs possible.

% \begin{figure}[t] 
% \centering
% \includegraphics[width=0.99\columnwidth]{Fig.5.Conductance_vs_B.png}
% \caption{The conductance oscillation of ALTJs of (a) even-layer ($N_z=10$) and (b) odd-layer ($N_z=11$) films at $E_F=30$ meV.
% }\label{Fig5}
% \end{figure}

\begin{acknowledgments}
This work was supported by the W\"urzburg-Dresden Cluster of Excellence on Complexity and Topology in Quantum Matter (EXC2147, project-id 390858490) and by the DFG (SFB1170 ``ToCoTronics''). We thank the Bavarian Ministry of Economic Affairs, Regional Development and Energy for financial support within the High-Tech Agenda Project ``Bausteine f\"ur das Quanten Computing auf Basis topologischer Materialen''.
\end{acknowledgments}

\begin{appendix}

\section{Derivation of the scattering matrix $S_{tb}$} \label{Sec:Appendix}
We adopt a matrix approach for the wavefunction matching method, which yields the scattering matrices $S_{tb}$. Since top and bottom surface states in the junction region are degenerate, four wavefunctions $\Psi_1^+(x),\, \Psi_1^-(x),\, \Psi_2^+(x),\, \Psi_2^-(x)$ are provided to be matched with incoming and outgoing scattering states $\Phi_t^+(x),\, \Phi_t^-(x),\, \Phi_b^+(x),\, \Phi_b^-(x)$ at $x=0$ and $x=L_x$. The wavefunction matching can be written in a matrix form as
\begin{eqnarray}
&&\left(\begin{array}{cccc}
\Phi_t^{+,in} & \mathbf{0} & \Phi_b^{+,in} & \textbf{0} \\
\mathbf{0} & \Phi_t^{-,in} & \mathbf{0} & \Phi_b^{-,in}
\end{array}\right)
= \left(\begin{array}{cc}
M & N
\end{array}\right)
\left(\begin{array}{c}
Q \nonumber\\
S_{tb}
\end{array}\right),\\
&&\text{where}\nonumber\\
&& M=\left(\begin{array}{cccc}
\Psi_1^+(0) & \Psi_1^-(0) & \Psi_2^+(0) & \Psi_2^-(0) \\
\Psi_1^+(L_x) & \Psi_1^-(L_x) & \Psi_2^+(L_x) & \Psi_2^-(L_x)
\end{array}\right), \nonumber\\
&& N=-\left(\begin{array}{cccc}
\mathbf{0}     & \Phi_t^{-,out} & \textbf{0}     &  \Phi_b^{-,out} \\
\Phi_t^{+,out} & \mathbf{0}     & \Phi_b^{+,out} &  \mathbf{0}
\end{array}\right).
\end{eqnarray}
$Q$ is a matrix whose elements form linear combinations of wavefunctions in the junction region. If we are only interested in the scattering matrix, the precise form of $Q$ is irrelevant. 
Our matrix approach yields the scattering matrix as

\begin{equation}
S_{tb} = \left(\begin{array}{cc} \mathbf{0} & \mathbf{1} \end{array}\right)
\left(\begin{array}{cc}
M & N
\end{array}\right)^{-1}
\left(\begin{array}{cccc}
\Phi_t^{+,in} & \mathbf{0} & \Phi_b^{+,in} & \textbf{0} \\
\mathbf{0} & \Phi_t^{-,in} & \mathbf{0} & \Phi_b^{-,in}
\end{array}\right).
\end{equation}

We apply the matrix approach to the derivation of the scattering matrix $S_{tb}^\text{odd}$ of odd layers. The scattering amplitudes of $S_{tb}$ are obtained as
\begin{eqnarray}
t'_\text{o} &=& \frac{1}{2}\sum_{s=\pm1}\frac{\hbar v_F k_s}{\hbar v_F k_s\cos(k_sL_x)-iE_F\sin(k_sL_x)}, \\
r'_\text{o} &=& \frac{-i}{2}\sum_{s=\pm1}\frac{ (m_\text{o} + s m_{tb})\sin(k_sL_x)}{\hbar v_F k_s\cos(k_sL_x)-iE_F\sin(k_sL_x)}, \\
t''_\text{o} &=& \frac{1}{2}\sum_{s=\pm1}\frac{s\hbar v_F k_s}{k_s\cos(k_sL_x)-iE_F\sin(k_sL_x)}, \\
r''_\text{o} &=& \frac{-i}{2}\sum_{s=\pm1}\frac{ (sm_\text{o} + m_{tb})\sin(k_sL_x)}{\hbar v_F k_s\cos(k_sL_x)-iE_F\sin(k_sL_x)},
\end{eqnarray}
with $k_\pm = \sqrt{E_F^2 - (m_\text{o}\pm m_{tb})^2}/(\hbar v_F)$. 
In addition, the scattering amplitudes of $S_{tb}^\text{even}$ for even-layers are obtained as
\begin{eqnarray}
t'_\text{e} &=& \frac{\hbar v_F k_F}{\hbar v_F k_F \cos{k_F L_x} - i E_F \sin(k_F L_x)}, \\
r'_\text{e} &=& \frac{-i m_\text{e}\sin(k_F L_x)}{\hbar v_F k_F \cos{k_F L_x} - i E_F \sin(k_F L_x)}, \\
r''_\text{e} &=& \frac{-i m_{tb}\sin(k_F L_x)}{\hbar v_F k_F \cos{k_F L_x} - i E_F \sin(k_F L_x)},
\end{eqnarray}
with $k_F = \sqrt{E_F^2 - m_\text{e}^2 - m_{tb}^2}/(\hbar v_F)$.

\end{appendix}

%%\bibliographystyle{apsrev4}
%\bibliography{Library-Sun}

\begin{thebibliography}{46}%
	\makeatletter
	\providecommand \@ifxundefined [1]{%
		\@ifx{#1\undefined}
	}%
	\providecommand \@ifnum [1]{%
		\ifnum #1\expandafter \@firstoftwo
		\else \expandafter \@secondoftwo
		\fi
	}%
	\providecommand \@ifx [1]{%
		\ifx #1\expandafter \@firstoftwo
		\else \expandafter \@secondoftwo
		\fi
	}%
	\providecommand \natexlab [1]{#1}%
	\providecommand \enquote  [1]{``#1''}%
	\providecommand \bibnamefont  [1]{#1}%
	\providecommand \bibfnamefont [1]{#1}%
	\providecommand \citenamefont [1]{#1}%
	\providecommand \href@noop [0]{\@secondoftwo}%
	\providecommand \href [0]{\begingroup \@sanitize@url \@href}%
	\providecommand \@href[1]{\@@startlink{#1}\@@href}%
	\providecommand \@@href[1]{\endgroup#1\@@endlink}%
	\providecommand \@sanitize@url [0]{\catcode `\\12\catcode `\$12\catcode
		`\&12\catcode `\#12\catcode `\^12\catcode `\_12\catcode `\%12\relax}%
	\providecommand \@@startlink[1]{}%
	\providecommand \@@endlink[0]{}%
	\providecommand \url  [0]{\begingroup\@sanitize@url \@url }%
	\providecommand \@url [1]{\endgroup\@href {#1}{\urlprefix }}%
	\providecommand \urlprefix  [0]{URL }%
	\providecommand \Eprint [0]{\href }%
	\providecommand \doibase [0]{https://doi.org/}%
	\providecommand \selectlanguage [0]{\@gobble}%
	\providecommand \bibinfo  [0]{\@secondoftwo}%
	\providecommand \bibfield  [0]{\@secondoftwo}%
	\providecommand \translation [1]{[#1]}%
	\providecommand \BibitemOpen [0]{}%
	\providecommand \bibitemStop [0]{}%
	\providecommand \bibitemNoStop [0]{.\EOS\space}%
	\providecommand \EOS [0]{\spacefactor3000\relax}%
	\providecommand \BibitemShut  [1]{\csname bibitem#1\endcsname}%
	\let\auto@bib@innerbib\@empty
	%</preamble>
	\bibitem [{\citenamefont {Liu}\ \emph {et~al.}(2016)\citenamefont {Liu},
		\citenamefont {Zhang},\ and\ \citenamefont {Qi}}]{Liu16ARCMPQuantum}%
	\BibitemOpen
	\bibfield  {author} {\bibinfo {author} {\bibfnamefont {C.-X.}\ \bibnamefont
			{Liu}}, \bibinfo {author} {\bibfnamefont {S.-C.}\ \bibnamefont {Zhang}},\
		and\ \bibinfo {author} {\bibfnamefont {X.-L.}\ \bibnamefont {Qi}},\
	}\bibfield  {title} {\bibinfo {title} {The {{Quantum Anomalous Hall Effect}}:
			{{Theory}} and {{Experiment}}},\ }\href
	{https://doi.org/10.1146/annurev-conmatphys-031115-011417} {\bibfield
		{journal} {\bibinfo  {journal} {Annual Review of Condensed Matter Physics}\
		}\textbf {\bibinfo {volume} {7}},\ \bibinfo {pages} {301} (\bibinfo {year}
		{2016})}\BibitemShut {NoStop}%
	\bibitem [{\citenamefont {Tokura}\ \emph {et~al.}(2019)\citenamefont {Tokura},
		\citenamefont {Yasuda},\ and\ \citenamefont
		{Tsukazaki}}]{Tokura19NRPMagnetic}%
	\BibitemOpen
	\bibfield  {author} {\bibinfo {author} {\bibfnamefont {Y.}~\bibnamefont
			{Tokura}}, \bibinfo {author} {\bibfnamefont {K.}~\bibnamefont {Yasuda}},\
		and\ \bibinfo {author} {\bibfnamefont {A.}~\bibnamefont {Tsukazaki}},\
	}\bibfield  {title} {\bibinfo {title} {Magnetic topological insulators},\
	}\href {https://www.nature.com/articles/s42254-018-0011-5} {\bibfield
		{journal} {\bibinfo  {journal} {Nat. Rev. Phys.}\ }\textbf {\bibinfo {volume}
			{1}},\ \bibinfo {pages} {126} (\bibinfo {year} {2019})}\BibitemShut {NoStop}%
	\bibitem [{\citenamefont {Wang}\ \emph {et~al.}(2021)\citenamefont {Wang},
		\citenamefont {Ge}, \citenamefont {Li}, \citenamefont {Liu}, \citenamefont
		{Xu},\ and\ \citenamefont {Wang}}]{Wang21IIntrinsic}%
	\BibitemOpen
	\bibfield  {author} {\bibinfo {author} {\bibfnamefont {P.}~\bibnamefont
			{Wang}}, \bibinfo {author} {\bibfnamefont {J.}~\bibnamefont {Ge}}, \bibinfo
		{author} {\bibfnamefont {J.}~\bibnamefont {Li}}, \bibinfo {author}
		{\bibfnamefont {Y.}~\bibnamefont {Liu}}, \bibinfo {author} {\bibfnamefont
			{Y.}~\bibnamefont {Xu}},\ and\ \bibinfo {author} {\bibfnamefont
			{J.}~\bibnamefont {Wang}},\ }\bibfield  {title} {\bibinfo {title} {Intrinsic
			magnetic topological insulators},\ }\href
	{https://www.sciencedirect.com/science/article/pii/S2666675821000230?via%3Dihub}
	{\bibfield  {journal} {\bibinfo  {journal} {Innovation}\ }\textbf {\bibinfo
			{volume} {2}} (\bibinfo {year} {2021})}\BibitemShut {NoStop}%
	\bibitem [{\citenamefont {Bernevig}\ \emph {et~al.}(2022)\citenamefont
		{Bernevig}, \citenamefont {Felser},\ and\ \citenamefont
		{Beidenkopf}}]{Bernevig22NProgress}%
	\BibitemOpen
	\bibfield  {author} {\bibinfo {author} {\bibfnamefont {B.~A.}\ \bibnamefont
			{Bernevig}}, \bibinfo {author} {\bibfnamefont {C.}~\bibnamefont {Felser}},\
		and\ \bibinfo {author} {\bibfnamefont {H.}~\bibnamefont {Beidenkopf}},\
	}\bibfield  {title} {\bibinfo {title} {Progress and prospects in magnetic
			topological materials},\ }\href {https://doi.org/10.1038/s41586-021-04105-x}
	{\bibfield  {journal} {\bibinfo  {journal} {Nature}\ }\textbf {\bibinfo
			{volume} {603}},\ \bibinfo {pages} {41} (\bibinfo {year} {2022})}\BibitemShut
	{NoStop}%
	\bibitem [{\citenamefont {Chang}\ \emph {et~al.}(2023)\citenamefont {Chang},
		\citenamefont {Liu},\ and\ \citenamefont {MacDonald}}]{Chang23RMPColloquium}%
	\BibitemOpen
	\bibfield  {author} {\bibinfo {author} {\bibfnamefont {C.-Z.}\ \bibnamefont
			{Chang}}, \bibinfo {author} {\bibfnamefont {C.-X.}\ \bibnamefont {Liu}},\
		and\ \bibinfo {author} {\bibfnamefont {A.~H.}\ \bibnamefont {MacDonald}},\
	}\bibfield  {title} {\bibinfo {title} {Colloquium: {{Quantum}} anomalous
			{{Hall}} effect},\ }\href {https://doi.org/10.1103/RevModPhys.95.011002}
	{\bibfield  {journal} {\bibinfo  {journal} {Rev. Mod. Phys.}\ }\textbf
		{\bibinfo {volume} {95}},\ \bibinfo {pages} {011002} (\bibinfo {year}
		{2023})}\BibitemShut {NoStop}%
	\bibitem [{\citenamefont {Wang}\ \emph {et~al.}(2023)\citenamefont {Wang},
		\citenamefont {Ma}, \citenamefont {Hao}, \citenamefont {Cai}, \citenamefont
		{Rong}, \citenamefont {Zhang}, \citenamefont {Chen}, \citenamefont {Zhang},
		\citenamefont {Lin}, \citenamefont {Zhao}, \citenamefont {Liu}, \citenamefont
		{Liu},\ and\ \citenamefont {Chen}}]{Wang23NSRTopological}%
	\BibitemOpen
	\bibfield  {author} {\bibinfo {author} {\bibfnamefont {Y.}~\bibnamefont
			{Wang}}, \bibinfo {author} {\bibfnamefont {X.-M.}\ \bibnamefont {Ma}},
		\bibinfo {author} {\bibfnamefont {Z.}~\bibnamefont {Hao}}, \bibinfo {author}
		{\bibfnamefont {Y.}~\bibnamefont {Cai}}, \bibinfo {author} {\bibfnamefont
			{H.}~\bibnamefont {Rong}}, \bibinfo {author} {\bibfnamefont {F.}~\bibnamefont
			{Zhang}}, \bibinfo {author} {\bibfnamefont {W.}~\bibnamefont {Chen}},
		\bibinfo {author} {\bibfnamefont {C.}~\bibnamefont {Zhang}}, \bibinfo
		{author} {\bibfnamefont {J.}~\bibnamefont {Lin}}, \bibinfo {author}
		{\bibfnamefont {Y.}~\bibnamefont {Zhao}}, \bibinfo {author} {\bibfnamefont
			{C.}~\bibnamefont {Liu}}, \bibinfo {author} {\bibfnamefont {Q.}~\bibnamefont
			{Liu}},\ and\ \bibinfo {author} {\bibfnamefont {C.}~\bibnamefont {Chen}},\
	}\bibfield  {title} {\bibinfo {title} {On the topological surface states of
			the intrinsic magnetic topological insulator {{Mn-Bi-Te}} family},\ }\href
	{https://doi.org/10.1093/nsr/nwad066} {\bibfield  {journal} {\bibinfo
			{journal} {National Science Review}\ ,\ \bibinfo {pages} {nwad066}} (\bibinfo
		{year} {2023})}\BibitemShut {NoStop}%
	\bibitem [{\citenamefont {Li}\ \emph {et~al.}(2023)\citenamefont {Li},
		\citenamefont {Liu}, \citenamefont {Liu}, \citenamefont {Wang}, \citenamefont
		{Lu},\ and\ \citenamefont {Xie}}]{Li23NSRProgress}%
	\BibitemOpen
	\bibfield  {author} {\bibinfo {author} {\bibfnamefont {S.}~\bibnamefont
			{Li}}, \bibinfo {author} {\bibfnamefont {T.}~\bibnamefont {Liu}}, \bibinfo
		{author} {\bibfnamefont {C.}~\bibnamefont {Liu}}, \bibinfo {author}
		{\bibfnamefont {Y.}~\bibnamefont {Wang}}, \bibinfo {author} {\bibfnamefont
			{H.-Z.}\ \bibnamefont {Lu}},\ and\ \bibinfo {author} {\bibfnamefont {X.~C.}\
			\bibnamefont {Xie}},\ }\bibfield  {title} {\bibinfo {title} {Progress on
			antiferromagnetic topological insulator {{MnBi2Te4}}},\ }\href
	{https://doi.org/10.1093/nsr/nwac296} {\bibfield  {journal} {\bibinfo
			{journal} {National Science Review}\ ,\ \bibinfo {pages} {nwac296}} (\bibinfo
		{year} {2023})}\BibitemShut {NoStop}%
	\bibitem [{\citenamefont {He}\ \emph {et~al.}(2014)\citenamefont {He},
		\citenamefont {Wang},\ and\ \citenamefont {Xue}}]{He14NSRQuantum}%
	\BibitemOpen
	\bibfield  {author} {\bibinfo {author} {\bibfnamefont {K.}~\bibnamefont
			{He}}, \bibinfo {author} {\bibfnamefont {Y.}~\bibnamefont {Wang}},\ and\
		\bibinfo {author} {\bibfnamefont {Q.-K.}\ \bibnamefont {Xue}},\ }\bibfield
	{title} {\bibinfo {title} {Quantum anomalous {{Hall}} effect},\ }\href
	{https://doi.org/10.1093/nsr/nwt029} {\bibfield  {journal} {\bibinfo
			{journal} {National Science Review}\ }\textbf {\bibinfo {volume} {1}},\
		\bibinfo {pages} {38} (\bibinfo {year} {2014})}\BibitemShut {NoStop}%
	\bibitem [{\citenamefont {Bhattacharyya}\ \emph {et~al.}(2021)\citenamefont
		{Bhattacharyya}, \citenamefont {Akhgar}, \citenamefont {Gebert},
		\citenamefont {Karel}, \citenamefont {Edmonds},\ and\ \citenamefont
		{Fuhrer}}]{Bhattacharyya21AMRecent}%
	\BibitemOpen
	\bibfield  {author} {\bibinfo {author} {\bibfnamefont {S.}~\bibnamefont
			{Bhattacharyya}}, \bibinfo {author} {\bibfnamefont {G.}~\bibnamefont
			{Akhgar}}, \bibinfo {author} {\bibfnamefont {M.}~\bibnamefont {Gebert}},
		\bibinfo {author} {\bibfnamefont {J.}~\bibnamefont {Karel}}, \bibinfo
		{author} {\bibfnamefont {M.~T.}\ \bibnamefont {Edmonds}},\ and\ \bibinfo
		{author} {\bibfnamefont {M.~S.}\ \bibnamefont {Fuhrer}},\ }\bibfield  {title}
	{\bibinfo {title} {Recent {{Progress}} in {{Proximity Coupling}} of
			{{Magnetism}} to {{Topological Insulators}}},\ }\href
	{https://doi.org/10.1002/adma.202007795} {\bibfield  {journal} {\bibinfo
			{journal} {Advanced Materials}\ }\textbf {\bibinfo {volume} {33}},\ \bibinfo
		{pages} {2007795} (\bibinfo {year} {2021})}\BibitemShut {NoStop}%
	\bibitem [{\citenamefont {Chi}\ and\ \citenamefont
		{Moodera}(2022)}]{Chi22AMProgress}%
	\BibitemOpen
	\bibfield  {author} {\bibinfo {author} {\bibfnamefont {H.}~\bibnamefont
			{Chi}}\ and\ \bibinfo {author} {\bibfnamefont {J.~S.}\ \bibnamefont
			{Moodera}},\ }\bibfield  {title} {\bibinfo {title} {Progress and prospects in
			the quantum anomalous {{Hall}} effect},\ }\href
	{https://doi.org/10.1063/5.0100989} {\bibfield  {journal} {\bibinfo
			{journal} {APL Materials}\ }\textbf {\bibinfo {volume} {10}},\ \bibinfo
		{pages} {090903} (\bibinfo {year} {2022})}\BibitemShut {NoStop}%
	\bibitem [{\citenamefont {Otrokov}\ \emph
		{et~al.}(2019{\natexlab{a}})\citenamefont {Otrokov}, \citenamefont
		{Klimovskikh}, \citenamefont {Bentmann}, \citenamefont {Estyunin},
		\citenamefont {Zeugner}, \citenamefont {Aliev}, \citenamefont {Ga{\ss}},
		\citenamefont {Wolter}, \citenamefont {Koroleva}, \citenamefont {Shikin},
		\citenamefont {{Blanco-Rey}}, \citenamefont {Hoffmann}, \citenamefont
		{Rusinov}, \citenamefont {Vyazovskaya}, \citenamefont {Eremeev},
		\citenamefont {Koroteev}, \citenamefont {Kuznetsov}, \citenamefont {Freyse},
		\citenamefont {{S{\'a}nchez-Barriga}}, \citenamefont {Amiraslanov},
		\citenamefont {Babanly}, \citenamefont {Mamedov}, \citenamefont {Abdullayev},
		\citenamefont {Zverev}, \citenamefont {Alfonsov}, \citenamefont {Kataev},
		\citenamefont {B{\"u}chner}, \citenamefont {Schwier}, \citenamefont {Kumar},
		\citenamefont {Kimura}, \citenamefont {Petaccia}, \citenamefont {Di~Santo},
		\citenamefont {Vidal}, \citenamefont {Schatz}, \citenamefont {Ki{\ss}ner},
		\citenamefont {{\"U}nzelmann}, \citenamefont {Min}, \citenamefont {Moser},
		\citenamefont {Peixoto}, \citenamefont {Reinert}, \citenamefont {Ernst},
		\citenamefont {Echenique}, \citenamefont {Isaeva},\ and\ \citenamefont
		{Chulkov}}]{Otrokov19NPrediction}%
	\BibitemOpen
	\bibfield  {author} {\bibinfo {author} {\bibfnamefont {M.~M.}\ \bibnamefont
			{Otrokov}}, \bibinfo {author} {\bibfnamefont {I.~I.}\ \bibnamefont
			{Klimovskikh}}, \bibinfo {author} {\bibfnamefont {H.}~\bibnamefont
			{Bentmann}}, \bibinfo {author} {\bibfnamefont {D.}~\bibnamefont {Estyunin}},
		\bibinfo {author} {\bibfnamefont {A.}~\bibnamefont {Zeugner}}, \bibinfo
		{author} {\bibfnamefont {Z.~S.}\ \bibnamefont {Aliev}}, \bibinfo {author}
		{\bibfnamefont {S.}~\bibnamefont {Ga{\ss}}}, \bibinfo {author} {\bibfnamefont
			{A.~U.~B.}\ \bibnamefont {Wolter}}, \bibinfo {author} {\bibfnamefont {A.~V.}\
			\bibnamefont {Koroleva}}, \bibinfo {author} {\bibfnamefont {A.~M.}\
			\bibnamefont {Shikin}}, \bibinfo {author} {\bibfnamefont {M.}~\bibnamefont
			{{Blanco-Rey}}}, \bibinfo {author} {\bibfnamefont {M.}~\bibnamefont
			{Hoffmann}}, \bibinfo {author} {\bibfnamefont {I.~P.}\ \bibnamefont
			{Rusinov}}, \bibinfo {author} {\bibfnamefont {A.~Y.}\ \bibnamefont
			{Vyazovskaya}}, \bibinfo {author} {\bibfnamefont {S.~V.}\ \bibnamefont
			{Eremeev}}, \bibinfo {author} {\bibfnamefont {Y.~M.}\ \bibnamefont
			{Koroteev}}, \bibinfo {author} {\bibfnamefont {V.~M.}\ \bibnamefont
			{Kuznetsov}}, \bibinfo {author} {\bibfnamefont {F.}~\bibnamefont {Freyse}},
		\bibinfo {author} {\bibfnamefont {J.}~\bibnamefont {{S{\'a}nchez-Barriga}}},
		\bibinfo {author} {\bibfnamefont {I.~R.}\ \bibnamefont {Amiraslanov}},
		\bibinfo {author} {\bibfnamefont {M.~B.}\ \bibnamefont {Babanly}}, \bibinfo
		{author} {\bibfnamefont {N.~T.}\ \bibnamefont {Mamedov}}, \bibinfo {author}
		{\bibfnamefont {N.~A.}\ \bibnamefont {Abdullayev}}, \bibinfo {author}
		{\bibfnamefont {V.~N.}\ \bibnamefont {Zverev}}, \bibinfo {author}
		{\bibfnamefont {A.}~\bibnamefont {Alfonsov}}, \bibinfo {author}
		{\bibfnamefont {V.}~\bibnamefont {Kataev}}, \bibinfo {author} {\bibfnamefont
			{B.}~\bibnamefont {B{\"u}chner}}, \bibinfo {author} {\bibfnamefont {E.~F.}\
			\bibnamefont {Schwier}}, \bibinfo {author} {\bibfnamefont {S.}~\bibnamefont
			{Kumar}}, \bibinfo {author} {\bibfnamefont {A.}~\bibnamefont {Kimura}},
		\bibinfo {author} {\bibfnamefont {L.}~\bibnamefont {Petaccia}}, \bibinfo
		{author} {\bibfnamefont {G.}~\bibnamefont {Di~Santo}}, \bibinfo {author}
		{\bibfnamefont {R.~C.}\ \bibnamefont {Vidal}}, \bibinfo {author}
		{\bibfnamefont {S.}~\bibnamefont {Schatz}}, \bibinfo {author} {\bibfnamefont
			{K.}~\bibnamefont {Ki{\ss}ner}}, \bibinfo {author} {\bibfnamefont
			{M.}~\bibnamefont {{\"U}nzelmann}}, \bibinfo {author} {\bibfnamefont {C.~H.}\
			\bibnamefont {Min}}, \bibinfo {author} {\bibfnamefont {S.}~\bibnamefont
			{Moser}}, \bibinfo {author} {\bibfnamefont {T.~R.~F.}\ \bibnamefont
			{Peixoto}}, \bibinfo {author} {\bibfnamefont {F.}~\bibnamefont {Reinert}},
		\bibinfo {author} {\bibfnamefont {A.}~\bibnamefont {Ernst}}, \bibinfo
		{author} {\bibfnamefont {P.~M.}\ \bibnamefont {Echenique}}, \bibinfo {author}
		{\bibfnamefont {A.}~\bibnamefont {Isaeva}},\ and\ \bibinfo {author}
		{\bibfnamefont {E.~V.}\ \bibnamefont {Chulkov}},\ }\bibfield  {title}
	{\bibinfo {title} {Prediction and observation of an antiferromagnetic
			topological insulator},\ }\href {https://doi.org/10.1038/s41586-019-1840-9}
	{\bibfield  {journal} {\bibinfo  {journal} {Nature}\ }\textbf {\bibinfo
			{volume} {576}},\ \bibinfo {pages} {416} (\bibinfo {year}
		{2019}{\natexlab{a}})}\BibitemShut {NoStop}%
	\bibitem [{\citenamefont {Deng}\ \emph {et~al.}(2020)\citenamefont {Deng},
		\citenamefont {Yu}, \citenamefont {Shi}, \citenamefont {Guo}, \citenamefont
		{Xu}, \citenamefont {Wang}, \citenamefont {Chen},\ and\ \citenamefont
		{Zhang}}]{Deng20SQuantum}%
	\BibitemOpen
	\bibfield  {author} {\bibinfo {author} {\bibfnamefont {Y.}~\bibnamefont
			{Deng}}, \bibinfo {author} {\bibfnamefont {Y.}~\bibnamefont {Yu}}, \bibinfo
		{author} {\bibfnamefont {M.~Z.}\ \bibnamefont {Shi}}, \bibinfo {author}
		{\bibfnamefont {Z.}~\bibnamefont {Guo}}, \bibinfo {author} {\bibfnamefont
			{Z.}~\bibnamefont {Xu}}, \bibinfo {author} {\bibfnamefont {J.}~\bibnamefont
			{Wang}}, \bibinfo {author} {\bibfnamefont {X.~H.}\ \bibnamefont {Chen}},\
		and\ \bibinfo {author} {\bibfnamefont {Y.}~\bibnamefont {Zhang}},\ }\bibfield
	{title} {\bibinfo {title} {Quantum anomalous {{Hall}} effect in intrinsic
			magnetic topological insulator {{MnBi}}$_2${{Te}}$_4$},\ }\href
	{https://science.sciencemag.org/content/367/6480/895} {\bibfield  {journal}
		{\bibinfo  {journal} {Science}\ }\textbf {\bibinfo {volume} {367}},\ \bibinfo
		{pages} {895} (\bibinfo {year} {2020})}\BibitemShut {NoStop}%
	\bibitem [{\citenamefont {Deng}\ \emph {et~al.}(2021)\citenamefont {Deng},
		\citenamefont {Chen}, \citenamefont {Wo{\l}o{\'s}}, \citenamefont
		{Konczykowski}, \citenamefont {Sobczak}, \citenamefont {Sitnicka},
		\citenamefont {Fedorchenko}, \citenamefont {Borysiuk}, \citenamefont
		{Heider}, \citenamefont {Pluci{\'n}ski}, \citenamefont {Park}, \citenamefont
		{Georgescu}, \citenamefont {Cano},\ and\ \citenamefont
		{{Krusin-Elbaum}}}]{Deng21NPHightemperature}%
	\BibitemOpen
	\bibfield  {author} {\bibinfo {author} {\bibfnamefont {H.}~\bibnamefont
			{Deng}}, \bibinfo {author} {\bibfnamefont {Z.}~\bibnamefont {Chen}}, \bibinfo
		{author} {\bibfnamefont {A.}~\bibnamefont {Wo{\l}o{\'s}}}, \bibinfo {author}
		{\bibfnamefont {M.}~\bibnamefont {Konczykowski}}, \bibinfo {author}
		{\bibfnamefont {K.}~\bibnamefont {Sobczak}}, \bibinfo {author} {\bibfnamefont
			{J.}~\bibnamefont {Sitnicka}}, \bibinfo {author} {\bibfnamefont {I.~V.}\
			\bibnamefont {Fedorchenko}}, \bibinfo {author} {\bibfnamefont
			{J.}~\bibnamefont {Borysiuk}}, \bibinfo {author} {\bibfnamefont
			{T.}~\bibnamefont {Heider}}, \bibinfo {author} {\bibfnamefont
			{{\L}.}~\bibnamefont {Pluci{\'n}ski}}, \bibinfo {author} {\bibfnamefont
			{K.}~\bibnamefont {Park}}, \bibinfo {author} {\bibfnamefont {A.~B.}\
			\bibnamefont {Georgescu}}, \bibinfo {author} {\bibfnamefont {J.}~\bibnamefont
			{Cano}},\ and\ \bibinfo {author} {\bibfnamefont {L.}~\bibnamefont
			{{Krusin-Elbaum}}},\ }\bibfield  {title} {\bibinfo {title} {High-temperature
			quantum anomalous {{Hall}} regime in a
			{{MnBi}}$_2${{Te}}$_4$/{{Bi}}$_2${{Te}}$_3$ superlattice},\ }\href
	{https://doi.org/10.1038/s41567-020-0998-2} {\bibfield  {journal} {\bibinfo
			{journal} {Nat. Phys.}\ }\textbf {\bibinfo {volume} {17}},\ \bibinfo {pages}
		{36} (\bibinfo {year} {2021})}\BibitemShut {NoStop}%
	\bibitem [{\citenamefont {Li}\ \emph {et~al.}(2019{\natexlab{a}})\citenamefont
		{Li}, \citenamefont {Li}, \citenamefont {Du}, \citenamefont {Wang},
		\citenamefont {Gu}, \citenamefont {Zhang}, \citenamefont {He}, \citenamefont
		{Duan},\ and\ \citenamefont {Xu}}]{Li19SAIntrinsic}%
	\BibitemOpen
	\bibfield  {author} {\bibinfo {author} {\bibfnamefont {J.}~\bibnamefont
			{Li}}, \bibinfo {author} {\bibfnamefont {Y.}~\bibnamefont {Li}}, \bibinfo
		{author} {\bibfnamefont {S.}~\bibnamefont {Du}}, \bibinfo {author}
		{\bibfnamefont {Z.}~\bibnamefont {Wang}}, \bibinfo {author} {\bibfnamefont
			{B.-L.}\ \bibnamefont {Gu}}, \bibinfo {author} {\bibfnamefont {S.-C.}\
			\bibnamefont {Zhang}}, \bibinfo {author} {\bibfnamefont {K.}~\bibnamefont
			{He}}, \bibinfo {author} {\bibfnamefont {W.}~\bibnamefont {Duan}},\ and\
		\bibinfo {author} {\bibfnamefont {Y.}~\bibnamefont {Xu}},\ }\bibfield
	{title} {\bibinfo {title} {Intrinsic magnetic topological insulators in van
			der {{Waals}} layered {{MnBi}}$_2${{Te}}$_4$-family materials},\ }\href
	{https://advances.sciencemag.org/content/5/6/eaaw5685} {\bibfield  {journal}
		{\bibinfo  {journal} {Sci. Adv.}\ }\textbf {\bibinfo {volume} {5}},\ \bibinfo
		{pages} {eaaw5685} (\bibinfo {year} {2019}{\natexlab{a}})}\BibitemShut
	{NoStop}%
	\bibitem [{\citenamefont {Li}\ \emph {et~al.}(2019{\natexlab{b}})\citenamefont
		{Li}, \citenamefont {Gao}, \citenamefont {Duan}, \citenamefont {Xu},
		\citenamefont {Zhu}, \citenamefont {Tian}, \citenamefont {Gao}, \citenamefont
		{Fan}, \citenamefont {Rao}, \citenamefont {Huang}, \citenamefont {Li},
		\citenamefont {Yan}, \citenamefont {Liu}, \citenamefont {Liu}, \citenamefont
		{Huang}, \citenamefont {Li}, \citenamefont {Liu}, \citenamefont {Zhang},
		\citenamefont {Zhang}, \citenamefont {Kondo}, \citenamefont {Shin},
		\citenamefont {Lei}, \citenamefont {Shi}, \citenamefont {Zhang},
		\citenamefont {Weng}, \citenamefont {Qian},\ and\ \citenamefont
		{Ding}}]{Li19PRXDirac}%
	\BibitemOpen
	\bibfield  {author} {\bibinfo {author} {\bibfnamefont {H.}~\bibnamefont
			{Li}}, \bibinfo {author} {\bibfnamefont {S.-Y.}\ \bibnamefont {Gao}},
		\bibinfo {author} {\bibfnamefont {S.-F.}\ \bibnamefont {Duan}}, \bibinfo
		{author} {\bibfnamefont {Y.-F.}\ \bibnamefont {Xu}}, \bibinfo {author}
		{\bibfnamefont {K.-J.}\ \bibnamefont {Zhu}}, \bibinfo {author} {\bibfnamefont
			{S.-J.}\ \bibnamefont {Tian}}, \bibinfo {author} {\bibfnamefont {J.-C.}\
			\bibnamefont {Gao}}, \bibinfo {author} {\bibfnamefont {W.-H.}\ \bibnamefont
			{Fan}}, \bibinfo {author} {\bibfnamefont {Z.-C.}\ \bibnamefont {Rao}},
		\bibinfo {author} {\bibfnamefont {J.-R.}\ \bibnamefont {Huang}}, \bibinfo
		{author} {\bibfnamefont {J.-J.}\ \bibnamefont {Li}}, \bibinfo {author}
		{\bibfnamefont {D.-Y.}\ \bibnamefont {Yan}}, \bibinfo {author} {\bibfnamefont
			{Z.-T.}\ \bibnamefont {Liu}}, \bibinfo {author} {\bibfnamefont {W.-L.}\
			\bibnamefont {Liu}}, \bibinfo {author} {\bibfnamefont {Y.-B.}\ \bibnamefont
			{Huang}}, \bibinfo {author} {\bibfnamefont {Y.-L.}\ \bibnamefont {Li}},
		\bibinfo {author} {\bibfnamefont {Y.}~\bibnamefont {Liu}}, \bibinfo {author}
		{\bibfnamefont {G.-B.}\ \bibnamefont {Zhang}}, \bibinfo {author}
		{\bibfnamefont {P.}~\bibnamefont {Zhang}}, \bibinfo {author} {\bibfnamefont
			{T.}~\bibnamefont {Kondo}}, \bibinfo {author} {\bibfnamefont
			{S.}~\bibnamefont {Shin}}, \bibinfo {author} {\bibfnamefont {H.-C.}\
			\bibnamefont {Lei}}, \bibinfo {author} {\bibfnamefont {Y.-G.}\ \bibnamefont
			{Shi}}, \bibinfo {author} {\bibfnamefont {W.-T.}\ \bibnamefont {Zhang}},
		\bibinfo {author} {\bibfnamefont {H.-M.}\ \bibnamefont {Weng}}, \bibinfo
		{author} {\bibfnamefont {T.}~\bibnamefont {Qian}},\ and\ \bibinfo {author}
		{\bibfnamefont {H.}~\bibnamefont {Ding}},\ }\bibfield  {title} {\bibinfo
		{title} {Dirac {{Surface States}} in {{Intrinsic Magnetic Topological
					Insulators EuSn}}$_2${{As}}$_2$ and {{MnBi}}$_{2n}${{Te}}$_{3n+1}$},\ }\href
	{https://link.aps.org/doi/10.1103/PhysRevX.9.041039} {\bibfield  {journal}
		{\bibinfo  {journal} {Phys. Rev. X}\ }\textbf {\bibinfo {volume} {9}},\
		\bibinfo {pages} {041039} (\bibinfo {year} {2019}{\natexlab{b}})}\BibitemShut
	{NoStop}%
	\bibitem [{\citenamefont {Hao}\ \emph {et~al.}(2019)\citenamefont {Hao},
		\citenamefont {Liu}, \citenamefont {Feng}, \citenamefont {Ma}, \citenamefont
		{Schwier}, \citenamefont {Arita}, \citenamefont {Kumar}, \citenamefont {Hu},
		\citenamefont {Lu}, \citenamefont {Zeng}, \citenamefont {Wang}, \citenamefont
		{Hao}, \citenamefont {Sun}, \citenamefont {Zhang}, \citenamefont {Mei},
		\citenamefont {Ni}, \citenamefont {Wu}, \citenamefont {Shimada},
		\citenamefont {Chen}, \citenamefont {Liu},\ and\ \citenamefont
		{Liu}}]{Hao19PRXGapless}%
	\BibitemOpen
	\bibfield  {author} {\bibinfo {author} {\bibfnamefont {Y.-J.}\ \bibnamefont
			{Hao}}, \bibinfo {author} {\bibfnamefont {P.}~\bibnamefont {Liu}}, \bibinfo
		{author} {\bibfnamefont {Y.}~\bibnamefont {Feng}}, \bibinfo {author}
		{\bibfnamefont {X.-M.}\ \bibnamefont {Ma}}, \bibinfo {author} {\bibfnamefont
			{E.~F.}\ \bibnamefont {Schwier}}, \bibinfo {author} {\bibfnamefont
			{M.}~\bibnamefont {Arita}}, \bibinfo {author} {\bibfnamefont
			{S.}~\bibnamefont {Kumar}}, \bibinfo {author} {\bibfnamefont
			{C.}~\bibnamefont {Hu}}, \bibinfo {author} {\bibfnamefont {R.}~\bibnamefont
			{Lu}}, \bibinfo {author} {\bibfnamefont {M.}~\bibnamefont {Zeng}}, \bibinfo
		{author} {\bibfnamefont {Y.}~\bibnamefont {Wang}}, \bibinfo {author}
		{\bibfnamefont {Z.}~\bibnamefont {Hao}}, \bibinfo {author} {\bibfnamefont
			{H.-Y.}\ \bibnamefont {Sun}}, \bibinfo {author} {\bibfnamefont
			{K.}~\bibnamefont {Zhang}}, \bibinfo {author} {\bibfnamefont
			{J.}~\bibnamefont {Mei}}, \bibinfo {author} {\bibfnamefont {N.}~\bibnamefont
			{Ni}}, \bibinfo {author} {\bibfnamefont {L.}~\bibnamefont {Wu}}, \bibinfo
		{author} {\bibfnamefont {K.}~\bibnamefont {Shimada}}, \bibinfo {author}
		{\bibfnamefont {C.}~\bibnamefont {Chen}}, \bibinfo {author} {\bibfnamefont
			{Q.}~\bibnamefont {Liu}},\ and\ \bibinfo {author} {\bibfnamefont
			{C.}~\bibnamefont {Liu}},\ }\bibfield  {title} {\bibinfo {title} {Gapless
			{{Surface Dirac Cone}} in {{Antiferromagnetic Topological Insulator
					MnBi}}$_2${{Te}}$_4$},\ }\href
	{https://link.aps.org/doi/10.1103/PhysRevX.9.041038} {\bibfield  {journal}
		{\bibinfo  {journal} {Phys. Rev. X}\ }\textbf {\bibinfo {volume} {9}},\
		\bibinfo {pages} {041038} (\bibinfo {year} {2019})}\BibitemShut {NoStop}%
	\bibitem [{\citenamefont {Zhang}\ \emph {et~al.}(2020)\citenamefont {Zhang},
		\citenamefont {Wu},\ and\ \citenamefont {Das~Sarma}}]{Zhang20PRLMobius}%
	\BibitemOpen
	\bibfield  {author} {\bibinfo {author} {\bibfnamefont {R.-X.}\ \bibnamefont
			{Zhang}}, \bibinfo {author} {\bibfnamefont {F.}~\bibnamefont {Wu}},\ and\
		\bibinfo {author} {\bibfnamefont {S.}~\bibnamefont {Das~Sarma}},\ }\bibfield
	{title} {\bibinfo {title} {M\"obius {{Insulator}} and {{Higher-Order
					Topology}} in {{MnBi}}$_{2n}${{Te}}$_{3n+1}$},\ }\href
	{https://doi.org/10.1103/PhysRevLett.124.136407} {\bibfield  {journal}
		{\bibinfo  {journal} {Phys. Rev. Lett.}\ }\textbf {\bibinfo {volume} {124}},\
		\bibinfo {pages} {136407} (\bibinfo {year} {2020})}\BibitemShut {NoStop}%
	\bibitem [{\citenamefont {Sun}\ \emph {et~al.}(2020)\citenamefont {Sun},
		\citenamefont {Wang}, \citenamefont {Zhang}, \citenamefont {Chen},
		\citenamefont {Zhao}, \citenamefont {Liu}, \citenamefont {Liu}, \citenamefont
		{Chen}, \citenamefont {Lu},\ and\ \citenamefont {Xie}}]{Sun20PRBAnalytical}%
	\BibitemOpen
	\bibfield  {author} {\bibinfo {author} {\bibfnamefont {H.-P.}\ \bibnamefont
			{Sun}}, \bibinfo {author} {\bibfnamefont {C.~M.}\ \bibnamefont {Wang}},
		\bibinfo {author} {\bibfnamefont {S.-B.}\ \bibnamefont {Zhang}}, \bibinfo
		{author} {\bibfnamefont {R.}~\bibnamefont {Chen}}, \bibinfo {author}
		{\bibfnamefont {Y.}~\bibnamefont {Zhao}}, \bibinfo {author} {\bibfnamefont
			{C.}~\bibnamefont {Liu}}, \bibinfo {author} {\bibfnamefont {Q.}~\bibnamefont
			{Liu}}, \bibinfo {author} {\bibfnamefont {C.}~\bibnamefont {Chen}}, \bibinfo
		{author} {\bibfnamefont {H.-Z.}\ \bibnamefont {Lu}},\ and\ \bibinfo {author}
		{\bibfnamefont {X.~C.}\ \bibnamefont {Xie}},\ }\bibfield  {title} {\bibinfo
		{title} {Analytical solution for the surface states of the antiferromagnetic
			topological insulator {{MnBi}}$_2${{Te}}$_4$},\ }\href
	{https://doi.org/10.1103/PhysRevB.102.241406} {\bibfield  {journal} {\bibinfo
			{journal} {Phys. Rev. B}\ }\textbf {\bibinfo {volume} {102}},\ \bibinfo
		{pages} {241406} (\bibinfo {year} {2020})}\BibitemShut {NoStop}%
	\bibitem [{\citenamefont {Wu}\ \emph {et~al.}(2020)\citenamefont {Wu},
		\citenamefont {Li}, \citenamefont {Ma}, \citenamefont {Zhang}, \citenamefont
		{Liu}, \citenamefont {Zhou}, \citenamefont {Shao}, \citenamefont {Wang},
		\citenamefont {Hao}, \citenamefont {Feng}, \citenamefont {Schwier},
		\citenamefont {Kumar}, \citenamefont {Sun}, \citenamefont {Liu},
		\citenamefont {Shimada}, \citenamefont {Miyamoto}, \citenamefont {Okuda},
		\citenamefont {Wang}, \citenamefont {Xie}, \citenamefont {Chen},
		\citenamefont {Liu}, \citenamefont {Liu},\ and\ \citenamefont
		{Zhao}}]{Wu20PRXDistinct}%
	\BibitemOpen
	\bibfield  {author} {\bibinfo {author} {\bibfnamefont {X.}~\bibnamefont
			{Wu}}, \bibinfo {author} {\bibfnamefont {J.}~\bibnamefont {Li}}, \bibinfo
		{author} {\bibfnamefont {X.-M.}\ \bibnamefont {Ma}}, \bibinfo {author}
		{\bibfnamefont {Y.}~\bibnamefont {Zhang}}, \bibinfo {author} {\bibfnamefont
			{Y.}~\bibnamefont {Liu}}, \bibinfo {author} {\bibfnamefont {C.-S.}\
			\bibnamefont {Zhou}}, \bibinfo {author} {\bibfnamefont {J.}~\bibnamefont
			{Shao}}, \bibinfo {author} {\bibfnamefont {Q.}~\bibnamefont {Wang}}, \bibinfo
		{author} {\bibfnamefont {Y.-J.}\ \bibnamefont {Hao}}, \bibinfo {author}
		{\bibfnamefont {Y.}~\bibnamefont {Feng}}, \bibinfo {author} {\bibfnamefont
			{E.~F.}\ \bibnamefont {Schwier}}, \bibinfo {author} {\bibfnamefont
			{S.}~\bibnamefont {Kumar}}, \bibinfo {author} {\bibfnamefont
			{H.}~\bibnamefont {Sun}}, \bibinfo {author} {\bibfnamefont {P.}~\bibnamefont
			{Liu}}, \bibinfo {author} {\bibfnamefont {K.}~\bibnamefont {Shimada}},
		\bibinfo {author} {\bibfnamefont {K.}~\bibnamefont {Miyamoto}}, \bibinfo
		{author} {\bibfnamefont {T.}~\bibnamefont {Okuda}}, \bibinfo {author}
		{\bibfnamefont {K.}~\bibnamefont {Wang}}, \bibinfo {author} {\bibfnamefont
			{M.}~\bibnamefont {Xie}}, \bibinfo {author} {\bibfnamefont {C.}~\bibnamefont
			{Chen}}, \bibinfo {author} {\bibfnamefont {Q.}~\bibnamefont {Liu}}, \bibinfo
		{author} {\bibfnamefont {C.}~\bibnamefont {Liu}},\ and\ \bibinfo {author}
		{\bibfnamefont {Y.}~\bibnamefont {Zhao}},\ }\bibfield  {title} {\bibinfo
		{title} {Distinct {{Topological Surface States}} on the {{Two Terminations}}
			of {{MnBi}}$_4${{Te}}$_7$},\ }\href
	{https://doi.org/10.1103/PhysRevX.10.031013} {\bibfield  {journal} {\bibinfo
			{journal} {Phys. Rev. X}\ }\textbf {\bibinfo {volume} {10}},\ \bibinfo
		{pages} {031013} (\bibinfo {year} {2020})}\BibitemShut {NoStop}%
	\bibitem [{\citenamefont {Yang}\ \emph {et~al.}(2021)\citenamefont {Yang},
		\citenamefont {Xu}, \citenamefont {Zhu}, \citenamefont {Niu}, \citenamefont
		{Xu}, \citenamefont {Peng}, \citenamefont {Cheng}, \citenamefont {Jia},
		\citenamefont {Huang}, \citenamefont {Xu}, \citenamefont {Lu},\ and\
		\citenamefont {Ye}}]{Yang21PRXOddEven}%
	\BibitemOpen
	\bibfield  {author} {\bibinfo {author} {\bibfnamefont {S.}~\bibnamefont
			{Yang}}, \bibinfo {author} {\bibfnamefont {X.}~\bibnamefont {Xu}}, \bibinfo
		{author} {\bibfnamefont {Y.}~\bibnamefont {Zhu}}, \bibinfo {author}
		{\bibfnamefont {R.}~\bibnamefont {Niu}}, \bibinfo {author} {\bibfnamefont
			{C.}~\bibnamefont {Xu}}, \bibinfo {author} {\bibfnamefont {Y.}~\bibnamefont
			{Peng}}, \bibinfo {author} {\bibfnamefont {X.}~\bibnamefont {Cheng}},
		\bibinfo {author} {\bibfnamefont {X.}~\bibnamefont {Jia}}, \bibinfo {author}
		{\bibfnamefont {Y.}~\bibnamefont {Huang}}, \bibinfo {author} {\bibfnamefont
			{X.}~\bibnamefont {Xu}}, \bibinfo {author} {\bibfnamefont {J.}~\bibnamefont
			{Lu}},\ and\ \bibinfo {author} {\bibfnamefont {Y.}~\bibnamefont {Ye}},\
	}\bibfield  {title} {\bibinfo {title} {Odd-{{Even Layer-Number Effect}} and
			{{Layer-Dependent Magnetic Phase Diagrams}} in {{MnBi}}$_{2}${{Te}}$_4$},\
	}\href {https://doi.org/10.1103/PhysRevX.11.011003} {\bibfield  {journal}
		{\bibinfo  {journal} {Phys. Rev. X}\ }\textbf {\bibinfo {volume} {11}},\
		\bibinfo {pages} {011003} (\bibinfo {year} {2021})}\BibitemShut {NoStop}%
	\bibitem [{\citenamefont {Li}\ \emph {et~al.}(2021)\citenamefont {Li},
		\citenamefont {Jiang}, \citenamefont {Chen},\ and\ \citenamefont
		{Xie}}]{Li21PRLCritical}%
	\BibitemOpen
	\bibfield  {author} {\bibinfo {author} {\bibfnamefont {H.}~\bibnamefont
			{Li}}, \bibinfo {author} {\bibfnamefont {H.}~\bibnamefont {Jiang}}, \bibinfo
		{author} {\bibfnamefont {C.-Z.}\ \bibnamefont {Chen}},\ and\ \bibinfo
		{author} {\bibfnamefont {X.~C.}\ \bibnamefont {Xie}},\ }\bibfield  {title}
	{\bibinfo {title} {Critical {{Behavior}} and {{Universal Signature}} of an
			{{Axion Insulator State}}},\ }\href
	{https://doi.org/10.1103/PhysRevLett.126.156601} {\bibfield  {journal}
		{\bibinfo  {journal} {Phys. Rev. Lett.}\ }\textbf {\bibinfo {volume} {126}},\
		\bibinfo {pages} {156601} (\bibinfo {year} {2021})}\BibitemShut {NoStop}%
	\bibitem [{\citenamefont {Chen}\ \emph {et~al.}(2023)\citenamefont {Chen},
		\citenamefont {Sun},\ and\ \citenamefont
		{Zhou}}]{Chen23PRBSidesurfacemediated}%
	\BibitemOpen
	\bibfield  {author} {\bibinfo {author} {\bibfnamefont {R.}~\bibnamefont
			{Chen}}, \bibinfo {author} {\bibfnamefont {H.-P.}\ \bibnamefont {Sun}},\ and\
		\bibinfo {author} {\bibfnamefont {B.}~\bibnamefont {Zhou}},\ }\bibfield
	{title} {\bibinfo {title} {Side-surface-mediated hybridization in axion
			insulators},\ }\href {https://doi.org/10.1103/PhysRevB.107.125304} {\bibfield
		{journal} {\bibinfo  {journal} {Phys. Rev. B}\ }\textbf {\bibinfo {volume}
			{107}},\ \bibinfo {pages} {125304} (\bibinfo {year} {2023})}\BibitemShut
	{NoStop}%
	\bibitem [{\citenamefont {Zhang}\ \emph {et~al.}(2019)\citenamefont {Zhang},
		\citenamefont {Shi}, \citenamefont {Zhu}, \citenamefont {Xing}, \citenamefont
		{Zhang},\ and\ \citenamefont {Wang}}]{Zhang19PRLTopological}%
	\BibitemOpen
	\bibfield  {author} {\bibinfo {author} {\bibfnamefont {D.}~\bibnamefont
			{Zhang}}, \bibinfo {author} {\bibfnamefont {M.}~\bibnamefont {Shi}}, \bibinfo
		{author} {\bibfnamefont {T.}~\bibnamefont {Zhu}}, \bibinfo {author}
		{\bibfnamefont {D.}~\bibnamefont {Xing}}, \bibinfo {author} {\bibfnamefont
			{H.}~\bibnamefont {Zhang}},\ and\ \bibinfo {author} {\bibfnamefont
			{J.}~\bibnamefont {Wang}},\ }\bibfield  {title} {\bibinfo {title}
		{Topological {{Axion States}} in the {{Magnetic Insulator
					MnBi}}$_2${{Te}}$_4$ with the {{Quantized Magnetoelectric Effect}}},\ }\href
	{https://link.aps.org/doi/10.1103/PhysRevLett.122.206401} {\bibfield
		{journal} {\bibinfo  {journal} {Phys. Rev. Lett.}\ }\textbf {\bibinfo
			{volume} {122}},\ \bibinfo {pages} {206401} (\bibinfo {year}
		{2019})}\BibitemShut {NoStop}%
	\bibitem [{\citenamefont {Otrokov}\ \emph
		{et~al.}(2019{\natexlab{b}})\citenamefont {Otrokov}, \citenamefont {Rusinov},
		\citenamefont {{Blanco-Rey}}, \citenamefont {Hoffmann}, \citenamefont
		{Vyazovskaya}, \citenamefont {Eremeev}, \citenamefont {Ernst}, \citenamefont
		{Echenique}, \citenamefont {Arnau},\ and\ \citenamefont
		{Chulkov}}]{Otrokov19PRLUnique}%
	\BibitemOpen
	\bibfield  {author} {\bibinfo {author} {\bibfnamefont {M.~M.}\ \bibnamefont
			{Otrokov}}, \bibinfo {author} {\bibfnamefont {I.~P.}\ \bibnamefont
			{Rusinov}}, \bibinfo {author} {\bibfnamefont {M.}~\bibnamefont
			{{Blanco-Rey}}}, \bibinfo {author} {\bibfnamefont {M.}~\bibnamefont
			{Hoffmann}}, \bibinfo {author} {\bibfnamefont {A.~{\relax Yu}.}\ \bibnamefont
			{Vyazovskaya}}, \bibinfo {author} {\bibfnamefont {S.~V.}\ \bibnamefont
			{Eremeev}}, \bibinfo {author} {\bibfnamefont {A.}~\bibnamefont {Ernst}},
		\bibinfo {author} {\bibfnamefont {P.~M.}\ \bibnamefont {Echenique}}, \bibinfo
		{author} {\bibfnamefont {A.}~\bibnamefont {Arnau}},\ and\ \bibinfo {author}
		{\bibfnamefont {E.~V.}\ \bibnamefont {Chulkov}},\ }\bibfield  {title}
	{\bibinfo {title} {Unique {{Thickness-Dependent Properties}} of the van der
			{{Waals Interlayer Antiferromagnet MnBi}}$_2${{Te}}$_4$ {{Films}}},\ }\href
	{https://link.aps.org/doi/10.1103/PhysRevLett.122.107202} {\bibfield
		{journal} {\bibinfo  {journal} {Phys. Rev. Lett.}\ }\textbf {\bibinfo
			{volume} {122}},\ \bibinfo {pages} {107202} (\bibinfo {year}
		{2019}{\natexlab{b}})}\BibitemShut {NoStop}%
	\bibitem [{\citenamefont {Liu}\ \emph {et~al.}(2020)\citenamefont {Liu},
		\citenamefont {Wang}, \citenamefont {Li}, \citenamefont {Wu}, \citenamefont
		{Li}, \citenamefont {Li}, \citenamefont {He}, \citenamefont {Xu},
		\citenamefont {Zhang},\ and\ \citenamefont {Wang}}]{Liu20NMRobust}%
	\BibitemOpen
	\bibfield  {author} {\bibinfo {author} {\bibfnamefont {C.}~\bibnamefont
			{Liu}}, \bibinfo {author} {\bibfnamefont {Y.}~\bibnamefont {Wang}}, \bibinfo
		{author} {\bibfnamefont {H.}~\bibnamefont {Li}}, \bibinfo {author}
		{\bibfnamefont {Y.}~\bibnamefont {Wu}}, \bibinfo {author} {\bibfnamefont
			{Y.}~\bibnamefont {Li}}, \bibinfo {author} {\bibfnamefont {J.}~\bibnamefont
			{Li}}, \bibinfo {author} {\bibfnamefont {K.}~\bibnamefont {He}}, \bibinfo
		{author} {\bibfnamefont {Y.}~\bibnamefont {Xu}}, \bibinfo {author}
		{\bibfnamefont {J.}~\bibnamefont {Zhang}},\ and\ \bibinfo {author}
		{\bibfnamefont {Y.}~\bibnamefont {Wang}},\ }\bibfield  {title} {\bibinfo
		{title} {Robust axion insulator and {{Chern}} insulator phases in a
			two-dimensional antiferromagnetic topological insulator},\ }\href
	{https://www.nature.com/articles/s41563-019-0573-3} {\bibfield  {journal}
		{\bibinfo  {journal} {Nat. Mater.}\ }\textbf {\bibinfo {volume} {19}},\
		\bibinfo {pages} {522} (\bibinfo {year} {2020})}\BibitemShut {NoStop}%
	\bibitem [{\citenamefont {Gao}\ \emph {et~al.}(2021)\citenamefont {Gao},
		\citenamefont {Liu}, \citenamefont {Hu}, \citenamefont {Qiu}, \citenamefont
		{Tzschaschel}, \citenamefont {Ghosh}, \citenamefont {Ho}, \citenamefont
		{B{\'e}rub{\'e}}, \citenamefont {Chen}, \citenamefont {Sun}, \citenamefont
		{Zhang}, \citenamefont {Zhang}, \citenamefont {Wang}, \citenamefont {Wang},
		\citenamefont {Huang}, \citenamefont {Felser}, \citenamefont {Agarwal},
		\citenamefont {Ding}, \citenamefont {Tien}, \citenamefont {Akey},
		\citenamefont {Gardener}, \citenamefont {Singh}, \citenamefont {Watanabe},
		\citenamefont {Taniguchi}, \citenamefont {Burch}, \citenamefont {Bell},
		\citenamefont {Zhou}, \citenamefont {Gao}, \citenamefont {Lu}, \citenamefont
		{Bansil}, \citenamefont {Lin}, \citenamefont {Chang}, \citenamefont {Fu},
		\citenamefont {Ma}, \citenamefont {Ni},\ and\ \citenamefont
		{Xu}}]{Gao21NLayer}%
	\BibitemOpen
	\bibfield  {author} {\bibinfo {author} {\bibfnamefont {A.}~\bibnamefont
			{Gao}}, \bibinfo {author} {\bibfnamefont {Y.-F.}\ \bibnamefont {Liu}},
		\bibinfo {author} {\bibfnamefont {C.}~\bibnamefont {Hu}}, \bibinfo {author}
		{\bibfnamefont {J.-X.}\ \bibnamefont {Qiu}}, \bibinfo {author} {\bibfnamefont
			{C.}~\bibnamefont {Tzschaschel}}, \bibinfo {author} {\bibfnamefont
			{B.}~\bibnamefont {Ghosh}}, \bibinfo {author} {\bibfnamefont {S.-C.}\
			\bibnamefont {Ho}}, \bibinfo {author} {\bibfnamefont {D.}~\bibnamefont
			{B{\'e}rub{\'e}}}, \bibinfo {author} {\bibfnamefont {R.}~\bibnamefont
			{Chen}}, \bibinfo {author} {\bibfnamefont {H.}~\bibnamefont {Sun}}, \bibinfo
		{author} {\bibfnamefont {Z.}~\bibnamefont {Zhang}}, \bibinfo {author}
		{\bibfnamefont {X.-Y.}\ \bibnamefont {Zhang}}, \bibinfo {author}
		{\bibfnamefont {Y.-X.}\ \bibnamefont {Wang}}, \bibinfo {author}
		{\bibfnamefont {N.}~\bibnamefont {Wang}}, \bibinfo {author} {\bibfnamefont
			{Z.}~\bibnamefont {Huang}}, \bibinfo {author} {\bibfnamefont
			{C.}~\bibnamefont {Felser}}, \bibinfo {author} {\bibfnamefont
			{A.}~\bibnamefont {Agarwal}}, \bibinfo {author} {\bibfnamefont
			{T.}~\bibnamefont {Ding}}, \bibinfo {author} {\bibfnamefont {H.-J.}\
			\bibnamefont {Tien}}, \bibinfo {author} {\bibfnamefont {A.}~\bibnamefont
			{Akey}}, \bibinfo {author} {\bibfnamefont {J.}~\bibnamefont {Gardener}},
		\bibinfo {author} {\bibfnamefont {B.}~\bibnamefont {Singh}}, \bibinfo
		{author} {\bibfnamefont {K.}~\bibnamefont {Watanabe}}, \bibinfo {author}
		{\bibfnamefont {T.}~\bibnamefont {Taniguchi}}, \bibinfo {author}
		{\bibfnamefont {K.~S.}\ \bibnamefont {Burch}}, \bibinfo {author}
		{\bibfnamefont {D.~C.}\ \bibnamefont {Bell}}, \bibinfo {author}
		{\bibfnamefont {B.~B.}\ \bibnamefont {Zhou}}, \bibinfo {author}
		{\bibfnamefont {W.}~\bibnamefont {Gao}}, \bibinfo {author} {\bibfnamefont
			{H.-Z.}\ \bibnamefont {Lu}}, \bibinfo {author} {\bibfnamefont
			{A.}~\bibnamefont {Bansil}}, \bibinfo {author} {\bibfnamefont
			{H.}~\bibnamefont {Lin}}, \bibinfo {author} {\bibfnamefont {T.-R.}\
			\bibnamefont {Chang}}, \bibinfo {author} {\bibfnamefont {L.}~\bibnamefont
			{Fu}}, \bibinfo {author} {\bibfnamefont {Q.}~\bibnamefont {Ma}}, \bibinfo
		{author} {\bibfnamefont {N.}~\bibnamefont {Ni}},\ and\ \bibinfo {author}
		{\bibfnamefont {S.-Y.}\ \bibnamefont {Xu}},\ }\bibfield  {title} {\bibinfo
		{title} {Layer {{Hall}} effect in a {{2D}} topological axion
			antiferromagnet},\ }\href {https://doi.org/10.1038/s41586-021-03679-w}
	{\bibfield  {journal} {\bibinfo  {journal} {Nature}\ }\textbf {\bibinfo
			{volume} {595}},\ \bibinfo {pages} {521} (\bibinfo {year}
		{2021})}\BibitemShut {NoStop}%
	\bibitem [{\citenamefont {Ge}\ \emph {et~al.}(2020)\citenamefont {Ge},
		\citenamefont {Liu}, \citenamefont {Li}, \citenamefont {Li}, \citenamefont
		{Luo}, \citenamefont {Wu}, \citenamefont {Xu},\ and\ \citenamefont
		{Wang}}]{Ge20NSRHighChernnumber}%
	\BibitemOpen
	\bibfield  {author} {\bibinfo {author} {\bibfnamefont {J.}~\bibnamefont
			{Ge}}, \bibinfo {author} {\bibfnamefont {Y.}~\bibnamefont {Liu}}, \bibinfo
		{author} {\bibfnamefont {J.}~\bibnamefont {Li}}, \bibinfo {author}
		{\bibfnamefont {H.}~\bibnamefont {Li}}, \bibinfo {author} {\bibfnamefont
			{T.}~\bibnamefont {Luo}}, \bibinfo {author} {\bibfnamefont {Y.}~\bibnamefont
			{Wu}}, \bibinfo {author} {\bibfnamefont {Y.}~\bibnamefont {Xu}},\ and\
		\bibinfo {author} {\bibfnamefont {J.}~\bibnamefont {Wang}},\ }\bibfield
	{title} {\bibinfo {title} {High-{{Chern-number}} and high-temperature quantum
			{{Hall}} effect without {{Landau}} levels},\ }\href
	{https://doi.org/10.1093/nsr/nwaa089} {\bibfield  {journal} {\bibinfo
			{journal} {Natl. Sci. Rev.}\ }\textbf {\bibinfo {volume} {7}},\ \bibinfo
		{pages} {1280} (\bibinfo {year} {2020})}\BibitemShut {NoStop}%
	\bibitem [{\citenamefont {Lee}\ \emph {et~al.}(2021)\citenamefont {Lee},
		\citenamefont {Graf}, \citenamefont {Min}, \citenamefont {Zhu}, \citenamefont
		{Yi}, \citenamefont {Ciocys}, \citenamefont {Wang}, \citenamefont {Choi},
		\citenamefont {Basnet}, \citenamefont {Fereidouni}, \citenamefont {Wegner},
		\citenamefont {Zhao}, \citenamefont {Verlinde}, \citenamefont {He},
		\citenamefont {Redwing}, \citenamefont {Gopalan}, \citenamefont {Churchill},
		\citenamefont {Lanzara}, \citenamefont {Samarth}, \citenamefont {Chang},
		\citenamefont {Hu},\ and\ \citenamefont {Mao}}]{Lee21PRXEvidence}%
	\BibitemOpen
	\bibfield  {author} {\bibinfo {author} {\bibfnamefont {S.~H.}\ \bibnamefont
			{Lee}}, \bibinfo {author} {\bibfnamefont {D.}~\bibnamefont {Graf}}, \bibinfo
		{author} {\bibfnamefont {L.}~\bibnamefont {Min}}, \bibinfo {author}
		{\bibfnamefont {Y.}~\bibnamefont {Zhu}}, \bibinfo {author} {\bibfnamefont
			{H.}~\bibnamefont {Yi}}, \bibinfo {author} {\bibfnamefont {S.}~\bibnamefont
			{Ciocys}}, \bibinfo {author} {\bibfnamefont {Y.}~\bibnamefont {Wang}},
		\bibinfo {author} {\bibfnamefont {E.~S.}\ \bibnamefont {Choi}}, \bibinfo
		{author} {\bibfnamefont {R.}~\bibnamefont {Basnet}}, \bibinfo {author}
		{\bibfnamefont {A.}~\bibnamefont {Fereidouni}}, \bibinfo {author}
		{\bibfnamefont {A.}~\bibnamefont {Wegner}}, \bibinfo {author} {\bibfnamefont
			{Y.-F.}\ \bibnamefont {Zhao}}, \bibinfo {author} {\bibfnamefont
			{K.}~\bibnamefont {Verlinde}}, \bibinfo {author} {\bibfnamefont
			{J.}~\bibnamefont {He}}, \bibinfo {author} {\bibfnamefont {R.}~\bibnamefont
			{Redwing}}, \bibinfo {author} {\bibfnamefont {V.}~\bibnamefont {Gopalan}},
		\bibinfo {author} {\bibfnamefont {H.~O.~H.}\ \bibnamefont {Churchill}},
		\bibinfo {author} {\bibfnamefont {A.}~\bibnamefont {Lanzara}}, \bibinfo
		{author} {\bibfnamefont {N.}~\bibnamefont {Samarth}}, \bibinfo {author}
		{\bibfnamefont {C.-Z.}\ \bibnamefont {Chang}}, \bibinfo {author}
		{\bibfnamefont {J.}~\bibnamefont {Hu}},\ and\ \bibinfo {author}
		{\bibfnamefont {Z.~Q.}\ \bibnamefont {Mao}},\ }\bibfield  {title} {\bibinfo
		{title} {Evidence for a {{Magnetic-Field-Induced Ideal Type-II Weyl State}}
			in {{Antiferromagnetic Topological Insulator
					Mn}}({Bi}$_{1-x}${Sb}$_x$)$_2${{Te}}$_4$},\ }\href
	{https://doi.org/10.1103/PhysRevX.11.031032} {\bibfield  {journal} {\bibinfo
			{journal} {Phys. Rev. X}\ }\textbf {\bibinfo {volume} {11}},\ \bibinfo
		{pages} {031032} (\bibinfo {year} {2021})}\BibitemShut {NoStop}%
	\bibitem [{\citenamefont {Lei}\ \emph {et~al.}(2022)\citenamefont {Lei},
		\citenamefont {Zhou}, \citenamefont {Hao}, \citenamefont {Liu}, \citenamefont
		{Yang}, \citenamefont {Sun}, \citenamefont {Ma}, \citenamefont {Ma},
		\citenamefont {Wang}, \citenamefont {Lu}, \citenamefont {Mei}, \citenamefont
		{Wang},\ and\ \citenamefont {He}}]{Lei22PRBMagnetically}%
	\BibitemOpen
	\bibfield  {author} {\bibinfo {author} {\bibfnamefont {X.}~\bibnamefont
			{Lei}}, \bibinfo {author} {\bibfnamefont {L.}~\bibnamefont {Zhou}}, \bibinfo
		{author} {\bibfnamefont {Z.}~\bibnamefont {Hao}}, \bibinfo {author}
		{\bibfnamefont {H.}~\bibnamefont {Liu}}, \bibinfo {author} {\bibfnamefont
			{S.}~\bibnamefont {Yang}}, \bibinfo {author} {\bibfnamefont {H.}~\bibnamefont
			{Sun}}, \bibinfo {author} {\bibfnamefont {X.}~\bibnamefont {Ma}}, \bibinfo
		{author} {\bibfnamefont {C.}~\bibnamefont {Ma}}, \bibinfo {author}
		{\bibfnamefont {L.}~\bibnamefont {Wang}}, \bibinfo {author} {\bibfnamefont
			{H.-Z.}\ \bibnamefont {Lu}}, \bibinfo {author} {\bibfnamefont {J.-W.}\
			\bibnamefont {Mei}}, \bibinfo {author} {\bibfnamefont {J.}~\bibnamefont
			{Wang}},\ and\ \bibinfo {author} {\bibfnamefont {H.}~\bibnamefont {He}},\
	}\bibfield  {title} {\bibinfo {title} {Magnetically tunable {{Shubnikov--de
					Haas}} oscillations in {{Mn}}{{Bi}}$_{2}${{Te}}$_{4}$},\ }\href
	{https://doi.org/10.1103/PhysRevB.105.155402} {\bibfield  {journal} {\bibinfo
			{journal} {Phys. Rev. B}\ }\textbf {\bibinfo {volume} {105}},\ \bibinfo
		{pages} {155402} (\bibinfo {year} {2022})}\BibitemShut {NoStop}%
	\bibitem [{\citenamefont {Xu}\ \emph {et~al.}(2019)\citenamefont {Xu},
		\citenamefont {Jiang}, \citenamefont {Miotkowski}, \citenamefont {Biswas},\
		and\ \citenamefont {Chen}}]{Xu19PRLTuning}%
	\BibitemOpen
	\bibfield  {author} {\bibinfo {author} {\bibfnamefont {Y.}~\bibnamefont
			{Xu}}, \bibinfo {author} {\bibfnamefont {G.}~\bibnamefont {Jiang}}, \bibinfo
		{author} {\bibfnamefont {I.}~\bibnamefont {Miotkowski}}, \bibinfo {author}
		{\bibfnamefont {R.~R.}\ \bibnamefont {Biswas}},\ and\ \bibinfo {author}
		{\bibfnamefont {Y.~P.}\ \bibnamefont {Chen}},\ }\bibfield  {title} {\bibinfo
		{title} {Tuning {{Insulator-Semimetal Transitions}} in {{3D Topological
					Insulator}} thin {{Films}} by {{Intersurface Hybridization}} and {{In-Plane
					Magnetic Fields}}},\ }\href {https://doi.org/10.1103/PhysRevLett.123.207701}
	{\bibfield  {journal} {\bibinfo  {journal} {Phys. Rev. Lett.}\ }\textbf
		{\bibinfo {volume} {123}},\ \bibinfo {pages} {207701} (\bibinfo {year}
		{2019})}\BibitemShut {NoStop}%
	\bibitem [{\citenamefont {Chong}\ \emph {et~al.}(2022)\citenamefont {Chong},
		\citenamefont {Liu}, \citenamefont {Watanabe}, \citenamefont {Taniguchi},
		\citenamefont {Sparks}, \citenamefont {Liu},\ and\ \citenamefont
		{Deshpande}}]{Chong22NCEmergent}%
	\BibitemOpen
	\bibfield  {author} {\bibinfo {author} {\bibfnamefont {S.~K.}\ \bibnamefont
			{Chong}}, \bibinfo {author} {\bibfnamefont {L.}~\bibnamefont {Liu}}, \bibinfo
		{author} {\bibfnamefont {K.}~\bibnamefont {Watanabe}}, \bibinfo {author}
		{\bibfnamefont {T.}~\bibnamefont {Taniguchi}}, \bibinfo {author}
		{\bibfnamefont {T.~D.}\ \bibnamefont {Sparks}}, \bibinfo {author}
		{\bibfnamefont {F.}~\bibnamefont {Liu}},\ and\ \bibinfo {author}
		{\bibfnamefont {V.~V.}\ \bibnamefont {Deshpande}},\ }\bibfield  {title}
	{\bibinfo {title} {Emergent helical edge states in a hybridized
			three-dimensional topological insulator},\ }\href
	{https://doi.org/10.1038/s41467-022-33643-9} {\bibfield  {journal} {\bibinfo
			{journal} {Nat Commun}\ }\textbf {\bibinfo {volume} {13}},\ \bibinfo {pages}
		{6386} (\bibinfo {year} {2022})}\BibitemShut {NoStop}%
	\bibitem [{\citenamefont {Michetti}\ and\ \citenamefont
		{Trauzettel}(2013)}]{Michetti13APLDevices}%
	\BibitemOpen
	\bibfield  {author} {\bibinfo {author} {\bibfnamefont {P.}~\bibnamefont
			{Michetti}}\ and\ \bibinfo {author} {\bibfnamefont {B.}~\bibnamefont
			{Trauzettel}},\ }\bibfield  {title} {\bibinfo {title} {Devices with
			electrically tunable topological insulating phases},\ }\href
	{https://doi.org/10.1063/1.4792275} {\bibfield  {journal} {\bibinfo
			{journal} {Appl. Phys. Lett.}\ }\textbf {\bibinfo {volume} {102}},\ \bibinfo
		{pages} {063503} (\bibinfo {year} {2013})}\BibitemShut {NoStop}%
	\bibitem [{\citenamefont {Qian}\ \emph {et~al.}(2014)\citenamefont {Qian},
		\citenamefont {Liu}, \citenamefont {Fu},\ and\ \citenamefont
		{Li}}]{Qian14SQuantum}%
	\BibitemOpen
	\bibfield  {author} {\bibinfo {author} {\bibfnamefont {X.}~\bibnamefont
			{Qian}}, \bibinfo {author} {\bibfnamefont {J.}~\bibnamefont {Liu}}, \bibinfo
		{author} {\bibfnamefont {L.}~\bibnamefont {Fu}},\ and\ \bibinfo {author}
		{\bibfnamefont {J.}~\bibnamefont {Li}},\ }\bibfield  {title} {\bibinfo
		{title} {Quantum spin {{Hall}} effect in two-dimensional transition metal
			dichalcogenides},\ }\href {https://doi.org/10.1126/science.1256815}
	{\bibfield  {journal} {\bibinfo  {journal} {Science}\ }\textbf {\bibinfo
			{volume} {346}},\ \bibinfo {pages} {1344} (\bibinfo {year}
		{2014})}\BibitemShut {NoStop}%
	\bibitem [{\citenamefont {Liu}\ \emph {et~al.}(2014)\citenamefont {Liu},
		\citenamefont {Hsieh}, \citenamefont {Wei}, \citenamefont {Duan},
		\citenamefont {Moodera},\ and\ \citenamefont {Fu}}]{Liu14NMSpinfiltered}%
	\BibitemOpen
	\bibfield  {author} {\bibinfo {author} {\bibfnamefont {J.}~\bibnamefont
			{Liu}}, \bibinfo {author} {\bibfnamefont {T.~H.}\ \bibnamefont {Hsieh}},
		\bibinfo {author} {\bibfnamefont {P.}~\bibnamefont {Wei}}, \bibinfo {author}
		{\bibfnamefont {W.}~\bibnamefont {Duan}}, \bibinfo {author} {\bibfnamefont
			{J.}~\bibnamefont {Moodera}},\ and\ \bibinfo {author} {\bibfnamefont
			{L.}~\bibnamefont {Fu}},\ }\bibfield  {title} {\bibinfo {title}
		{Spin-filtered edge states with an electrically tunable gap in a
			two-dimensional topological crystalline insulator},\ }\href
	{https://doi.org/10.1038/nmat3828} {\bibfield  {journal} {\bibinfo  {journal}
			{Nat. Mater.}\ }\textbf {\bibinfo {volume} {13}},\ \bibinfo {pages} {178}
		(\bibinfo {year} {2014})}\BibitemShut {NoStop}%
	\bibitem [{\citenamefont {Wang}\ \emph {et~al.}(2015)\citenamefont {Wang},
		\citenamefont {Lian},\ and\ \citenamefont {Zhang}}]{Wang15PRLElectrically}%
	\BibitemOpen
	\bibfield  {author} {\bibinfo {author} {\bibfnamefont {J.}~\bibnamefont
			{Wang}}, \bibinfo {author} {\bibfnamefont {B.}~\bibnamefont {Lian}},\ and\
		\bibinfo {author} {\bibfnamefont {S.-C.}\ \bibnamefont {Zhang}},\ }\bibfield
	{title} {\bibinfo {title} {Electrically {{Tunable Magnetism}} in {{Magnetic
					Topological Insulators}}},\ }\href
	{https://doi.org/10.1103/PhysRevLett.115.036805} {\bibfield  {journal}
		{\bibinfo  {journal} {Phys. Rev. Lett.}\ }\textbf {\bibinfo {volume} {115}},\
		\bibinfo {pages} {036805} (\bibinfo {year} {2015})}\BibitemShut {NoStop}%
	\bibitem [{\citenamefont {Liu}\ \emph {et~al.}(2015)\citenamefont {Liu},
		\citenamefont {Zhang}, \citenamefont {Abdalla}, \citenamefont {Fazzio},\ and\
		\citenamefont {Zunger}}]{Liu15NLSwitching}%
	\BibitemOpen
	\bibfield  {author} {\bibinfo {author} {\bibfnamefont {Q.}~\bibnamefont
			{Liu}}, \bibinfo {author} {\bibfnamefont {X.}~\bibnamefont {Zhang}}, \bibinfo
		{author} {\bibfnamefont {L.~B.}\ \bibnamefont {Abdalla}}, \bibinfo {author}
		{\bibfnamefont {A.}~\bibnamefont {Fazzio}},\ and\ \bibinfo {author}
		{\bibfnamefont {A.}~\bibnamefont {Zunger}},\ }\bibfield  {title} {\bibinfo
		{title} {Switching a {{Normal Insulator}} into a {{Topological Insulator}}
			via {{Electric Field}} with {{Application}} to {{Phosphorene}}},\ }\href
	{https://doi.org/10.1021/nl5043769} {\bibfield  {journal} {\bibinfo
			{journal} {Nano Lett.}\ }\textbf {\bibinfo {volume} {15}},\ \bibinfo {pages}
		{1222} (\bibinfo {year} {2015})}\BibitemShut {NoStop}%
	\bibitem [{\citenamefont {Collins}\ \emph {et~al.}(2018)\citenamefont
		{Collins}, \citenamefont {Tadich}, \citenamefont {Wu}, \citenamefont {Gomes},
		\citenamefont {Rodrigues}, \citenamefont {Liu}, \citenamefont {Hellerstedt},
		\citenamefont {Ryu}, \citenamefont {Tang}, \citenamefont {Mo}, \citenamefont
		{Adam}, \citenamefont {Yang}, \citenamefont {Fuhrer},\ and\ \citenamefont
		{Edmonds}}]{Collins18NElectricfieldtuned}%
	\BibitemOpen
	\bibfield  {author} {\bibinfo {author} {\bibfnamefont {J.~L.}\ \bibnamefont
			{Collins}}, \bibinfo {author} {\bibfnamefont {A.}~\bibnamefont {Tadich}},
		\bibinfo {author} {\bibfnamefont {W.}~\bibnamefont {Wu}}, \bibinfo {author}
		{\bibfnamefont {L.~C.}\ \bibnamefont {Gomes}}, \bibinfo {author}
		{\bibfnamefont {J.~N.~B.}\ \bibnamefont {Rodrigues}}, \bibinfo {author}
		{\bibfnamefont {C.}~\bibnamefont {Liu}}, \bibinfo {author} {\bibfnamefont
			{J.}~\bibnamefont {Hellerstedt}}, \bibinfo {author} {\bibfnamefont
			{H.}~\bibnamefont {Ryu}}, \bibinfo {author} {\bibfnamefont {S.}~\bibnamefont
			{Tang}}, \bibinfo {author} {\bibfnamefont {S.-K.}\ \bibnamefont {Mo}},
		\bibinfo {author} {\bibfnamefont {S.}~\bibnamefont {Adam}}, \bibinfo {author}
		{\bibfnamefont {S.~A.}\ \bibnamefont {Yang}}, \bibinfo {author}
		{\bibfnamefont {M.~S.}\ \bibnamefont {Fuhrer}},\ and\ \bibinfo {author}
		{\bibfnamefont {M.~T.}\ \bibnamefont {Edmonds}},\ }\bibfield  {title}
	{\bibinfo {title} {Electric-field-tuned topological phase transition in
			ultrathin {{Na3Bi}}},\ }\href {https://doi.org/10.1038/s41586-018-0788-5}
	{\bibfield  {journal} {\bibinfo  {journal} {Nature}\ }\textbf {\bibinfo
			{volume} {564}},\ \bibinfo {pages} {390} (\bibinfo {year}
		{2018})}\BibitemShut {NoStop}%
	\bibitem [{\citenamefont {Sun}\ \emph {et~al.}(2023)\citenamefont {Sun},
		\citenamefont {Li}, \citenamefont {Choi}, \citenamefont {Zhang},
		\citenamefont {Lu},\ and\ \citenamefont {Trauzettel}}]{Sun23PRRMagnetic}%
	\BibitemOpen
	\bibfield  {author} {\bibinfo {author} {\bibfnamefont {H.-P.}\ \bibnamefont
			{Sun}}, \bibinfo {author} {\bibfnamefont {C.-A.}\ \bibnamefont {Li}},
		\bibinfo {author} {\bibfnamefont {S.-J.}\ \bibnamefont {Choi}}, \bibinfo
		{author} {\bibfnamefont {S.-B.}\ \bibnamefont {Zhang}}, \bibinfo {author}
		{\bibfnamefont {H.-Z.}\ \bibnamefont {Lu}},\ and\ \bibinfo {author}
		{\bibfnamefont {B.}~\bibnamefont {Trauzettel}},\ }\bibfield  {title}
	{\bibinfo {title} {Magnetic topological transistor exploiting layer-selective
			transport},\ }\href {https://doi.org/10.1103/PhysRevResearch.5.013179}
	{\bibfield  {journal} {\bibinfo  {journal} {Phys. Rev. Res.}\ }\textbf
		{\bibinfo {volume} {5}},\ \bibinfo {pages} {013179} (\bibinfo {year}
		{2023})}\BibitemShut {NoStop}%
	\bibitem [{\citenamefont {Gong}\ \emph {et~al.}(2019)\citenamefont {Gong},
		\citenamefont {Guo}, \citenamefont {Li}, \citenamefont {Zhu}, \citenamefont
		{Liao}, \citenamefont {Liu}, \citenamefont {Zhang}, \citenamefont {Gu},
		\citenamefont {Tang}, \citenamefont {Feng}, \citenamefont {Zhang},
		\citenamefont {Li}, \citenamefont {Song}, \citenamefont {Wang}, \citenamefont
		{Yu}, \citenamefont {Chen}, \citenamefont {Wang}, \citenamefont {Yao},
		\citenamefont {Duan}, \citenamefont {Xu}, \citenamefont {Zhang},
		\citenamefont {Ma}, \citenamefont {Xue},\ and\ \citenamefont
		{He}}]{Gong19CPLExperimental}%
	\BibitemOpen
	\bibfield  {author} {\bibinfo {author} {\bibfnamefont {Y.}~\bibnamefont
			{Gong}}, \bibinfo {author} {\bibfnamefont {J.}~\bibnamefont {Guo}}, \bibinfo
		{author} {\bibfnamefont {J.}~\bibnamefont {Li}}, \bibinfo {author}
		{\bibfnamefont {K.}~\bibnamefont {Zhu}}, \bibinfo {author} {\bibfnamefont
			{M.}~\bibnamefont {Liao}}, \bibinfo {author} {\bibfnamefont {X.}~\bibnamefont
			{Liu}}, \bibinfo {author} {\bibfnamefont {Q.}~\bibnamefont {Zhang}}, \bibinfo
		{author} {\bibfnamefont {L.}~\bibnamefont {Gu}}, \bibinfo {author}
		{\bibfnamefont {L.}~\bibnamefont {Tang}}, \bibinfo {author} {\bibfnamefont
			{X.}~\bibnamefont {Feng}}, \bibinfo {author} {\bibfnamefont {D.}~\bibnamefont
			{Zhang}}, \bibinfo {author} {\bibfnamefont {W.}~\bibnamefont {Li}}, \bibinfo
		{author} {\bibfnamefont {C.}~\bibnamefont {Song}}, \bibinfo {author}
		{\bibfnamefont {L.}~\bibnamefont {Wang}}, \bibinfo {author} {\bibfnamefont
			{P.}~\bibnamefont {Yu}}, \bibinfo {author} {\bibfnamefont {X.}~\bibnamefont
			{Chen}}, \bibinfo {author} {\bibfnamefont {Y.}~\bibnamefont {Wang}}, \bibinfo
		{author} {\bibfnamefont {H.}~\bibnamefont {Yao}}, \bibinfo {author}
		{\bibfnamefont {W.}~\bibnamefont {Duan}}, \bibinfo {author} {\bibfnamefont
			{Y.}~\bibnamefont {Xu}}, \bibinfo {author} {\bibfnamefont {S.-C.}\
			\bibnamefont {Zhang}}, \bibinfo {author} {\bibfnamefont {X.}~\bibnamefont
			{Ma}}, \bibinfo {author} {\bibfnamefont {Q.-K.}\ \bibnamefont {Xue}},\ and\
		\bibinfo {author} {\bibfnamefont {K.}~\bibnamefont {He}},\ }\bibfield
	{title} {\bibinfo {title} {Experimental {{Realization}} of an {{Intrinsic
					Magnetic Topological Insulator}}},\ }\href
	{https://doi.org/10.1088/0256-307x/36/7/076801} {\bibfield  {journal}
		{\bibinfo  {journal} {Chinese Phys. Lett.}\ }\textbf {\bibinfo {volume}
			{36}},\ \bibinfo {pages} {076801} (\bibinfo {year} {2019})}\BibitemShut
	{NoStop}%
	\bibitem [{\citenamefont {Cai}\ \emph {et~al.}(2022)\citenamefont {Cai},
		\citenamefont {Ovchinnikov}, \citenamefont {Fei}, \citenamefont {He},
		\citenamefont {Song}, \citenamefont {Lin}, \citenamefont {Wang},
		\citenamefont {Cobden}, \citenamefont {Chu}, \citenamefont {Cui},
		\citenamefont {Chang}, \citenamefont {Xiao}, \citenamefont {Yan},\ and\
		\citenamefont {Xu}}]{Cai22NCElectric}%
	\BibitemOpen
	\bibfield  {author} {\bibinfo {author} {\bibfnamefont {J.}~\bibnamefont
			{Cai}}, \bibinfo {author} {\bibfnamefont {D.}~\bibnamefont {Ovchinnikov}},
		\bibinfo {author} {\bibfnamefont {Z.}~\bibnamefont {Fei}}, \bibinfo {author}
		{\bibfnamefont {M.}~\bibnamefont {He}}, \bibinfo {author} {\bibfnamefont
			{T.}~\bibnamefont {Song}}, \bibinfo {author} {\bibfnamefont {Z.}~\bibnamefont
			{Lin}}, \bibinfo {author} {\bibfnamefont {C.}~\bibnamefont {Wang}}, \bibinfo
		{author} {\bibfnamefont {D.}~\bibnamefont {Cobden}}, \bibinfo {author}
		{\bibfnamefont {J.-H.}\ \bibnamefont {Chu}}, \bibinfo {author} {\bibfnamefont
			{Y.-T.}\ \bibnamefont {Cui}}, \bibinfo {author} {\bibfnamefont {C.-Z.}\
			\bibnamefont {Chang}}, \bibinfo {author} {\bibfnamefont {D.}~\bibnamefont
			{Xiao}}, \bibinfo {author} {\bibfnamefont {J.}~\bibnamefont {Yan}},\ and\
		\bibinfo {author} {\bibfnamefont {X.}~\bibnamefont {Xu}},\ }\bibfield
	{title} {\bibinfo {title} {Electric control of a canted-antiferromagnetic
			{{Chern}} insulator},\ }\href {https://doi.org/10.1038/s41467-022-29259-8}
	{\bibfield  {journal} {\bibinfo  {journal} {Nat. Commun.}\ }\textbf {\bibinfo
			{volume} {13}},\ \bibinfo {pages} {1668} (\bibinfo {year}
		{2022})}\BibitemShut {NoStop}%
	\bibitem [{\citenamefont {Lu}\ \emph {et~al.}(2010)\citenamefont {Lu},
		\citenamefont {Shan}, \citenamefont {Yao}, \citenamefont {Niu},\ and\
		\citenamefont {Shen}}]{Lu10PRBMassive}%
	\BibitemOpen
	\bibfield  {author} {\bibinfo {author} {\bibfnamefont {H.-Z.}\ \bibnamefont
			{Lu}}, \bibinfo {author} {\bibfnamefont {W.-Y.}\ \bibnamefont {Shan}},
		\bibinfo {author} {\bibfnamefont {W.}~\bibnamefont {Yao}}, \bibinfo {author}
		{\bibfnamefont {Q.}~\bibnamefont {Niu}},\ and\ \bibinfo {author}
		{\bibfnamefont {S.-Q.}\ \bibnamefont {Shen}},\ }\bibfield  {title} {\bibinfo
		{title} {Massive {{Dirac}} fermions and spin physics in an ultrathin film of
			topological insulator},\ }\href {https://doi.org/10.1103/PhysRevB.81.115407}
	{\bibfield  {journal} {\bibinfo  {journal} {Phys. Rev. B}\ }\textbf {\bibinfo
			{volume} {81}},\ \bibinfo {pages} {115407} (\bibinfo {year}
		{2010})}\BibitemShut {NoStop}%
	\bibitem [{\citenamefont {Shan}\ \emph {et~al.}(2010)\citenamefont {Shan},
		\citenamefont {Lu},\ and\ \citenamefont {Shen}}]{Shan10NJPEffective}%
	\BibitemOpen
	\bibfield  {author} {\bibinfo {author} {\bibfnamefont {W.-Y.}\ \bibnamefont
			{Shan}}, \bibinfo {author} {\bibfnamefont {H.-Z.}\ \bibnamefont {Lu}},\ and\
		\bibinfo {author} {\bibfnamefont {S.-Q.}\ \bibnamefont {Shen}},\ }\bibfield
	{title} {\bibinfo {title} {Effective continuous model for surface states and
			thin films of three-dimensional topological insulators},\ }\href
	{https://doi.org/10.1088/1367-2630/12/4/043048} {\bibfield  {journal}
		{\bibinfo  {journal} {New J. Phys.}\ }\textbf {\bibinfo {volume} {12}},\
		\bibinfo {pages} {043048} (\bibinfo {year} {2010})}\BibitemShut {NoStop}%
	\bibitem [{\citenamefont {Landauer}(1970)}]{Landauer70Electrical}%
	\BibitemOpen
	\bibfield  {author} {\bibinfo {author} {\bibfnamefont {R.}~\bibnamefont
			{Landauer}},\ }\bibfield  {title} {\bibinfo {title} {Electrical resistance of
			disordered one-dimensional lattices},\ }\href
	{https://doi.org/10.1080/14786437008238472} {\bibfield  {journal} {\bibinfo
			{journal} {The Philosophical Magazine}\ }\textbf {\bibinfo {volume} {21}},\
		\bibinfo {pages} {863} (\bibinfo {year} {1970})}\BibitemShut {NoStop}%
	\bibitem [{\citenamefont {B{\"u}ttiker}(1986)}]{Buttiker86Four}%
	\BibitemOpen
	\bibfield  {author} {\bibinfo {author} {\bibfnamefont {M.}~\bibnamefont
			{B{\"u}ttiker}},\ }\bibfield  {title} {\bibinfo {title} {Four-{{Terminal
					Phase-Coherent Conductance}}},\ }\href
	{https://doi.org/10.1103/PhysRevLett.57.1761} {\bibfield  {journal} {\bibinfo
			{journal} {Phys. Rev. Lett.}\ }\textbf {\bibinfo {volume} {57}},\ \bibinfo
		{pages} {1761} (\bibinfo {year} {1986})}\BibitemShut {NoStop}%
	\bibitem [{\citenamefont {Datta}(1997)}]{Datta97Electronic}%
	\BibitemOpen
	\bibfield  {author} {\bibinfo {author} {\bibfnamefont {S.}~\bibnamefont
			{Datta}},\ }\href@noop {} {\emph {\bibinfo {title} {Electronic {{Transport}}
				in {{Mesoscopic Systems}}}}}\ (\bibinfo  {publisher} {{Cambridge University
			Press}},\ \bibinfo {year} {1997})\BibitemShut {NoStop}%
	\bibitem [{\citenamefont {Choi}\ \emph {et~al.}(2013)\citenamefont {Choi},
		\citenamefont {Park},\ and\ \citenamefont {Sim}}]{13Choi}%
	\BibitemOpen
	\bibfield  {author} {\bibinfo {author} {\bibfnamefont {S.-J.}\ \bibnamefont
			{Choi}}, \bibinfo {author} {\bibfnamefont {S.}~\bibnamefont {Park}},\ and\
		\bibinfo {author} {\bibfnamefont {H.-S.}\ \bibnamefont {Sim}},\ }\bibfield
	{title} {\bibinfo {title} {Tunable geometric phase of dirac fermions in a
			topological junction},\ }\href {https://doi.org/10.1103/PhysRevB.87.165420}
	{\bibfield  {journal} {\bibinfo  {journal} {Phys. Rev. B}\ }\textbf {\bibinfo
			{volume} {87}},\ \bibinfo {pages} {165420} (\bibinfo {year}
		{2013})}\BibitemShut {NoStop}%
\end{thebibliography}%


\end{document}